\documentclass[journal,peerreview]{IEEEtran}
\usepackage[utf8]{inputenc}
\usepackage{graphicx}
\usepackage{algorithmicx, algpseudocode, algorithm}
\usepackage{xcolor}
\usepackage[caption=false,font=footnotesize]{subfig}
\usepackage{amsmath, amssymb, amsthm, mathtools}

\newtheorem{property}{Property}
\newtheorem{definition}{Definition}
\newtheorem{myrule}{Rule}

\begin{document}

\title{TDSR: Transparent Distributed Segment-Based Routing}

\author{Juan-Jose~Crespo,
        German~Maglione-Mathey,
        Jos\'{e}~L.~S\'{a}nchez,
        Francisco~J.~Alfaro-Cort\'{e}s,
        Jose~Flich
\IEEEcompsocitemizethanks{
    \IEEEcompsocthanksitem{
        Juan-Jose~Crespo, German~Maglione-Mathey,
        Jos\'{e}~L.~S\'{a}nchez and
        Francisco~J.~Alfaro-Cort\'{e}s are with the Dept. of Computing
        Systems, University of Castilla-La Mancha, Albacete, Spain.

        Email: \{juanjose.gcrespo ; jose.sgarcia ; fco.alfaro
        \}@uclm.es, german.maglione@dsi.uclm.es
}

    \IEEEcompsocthanksitem{
        Jose~Flich is with the Dept. of Computer
        Engineering, Technical University of Valencia, Valencia, Spain.

        Email: jflich@disca.upv.es
}
}}


\maketitle

\begin{abstract}
Component reliability and performance pose a great challenge for
interconnection networks. Future technology scaling such as transistor
integration capacity in VLSI design will result in higher device
degradation and manufacture variability. As a consequence, changes in
the network arise, often rendering irregular topologies. This paper
proposes a topology-agnostic distributed segment-based algorithm able to
handle switch discovery in any topology while guaranteeing connectivity
among switches. The proposal, known as Transparent Distributed
Segment-Based Routing (TDSR), has been applied to meshes with defective
link configurations.
\end{abstract}

\begin{IEEEkeywords}
topology-agnostic, distributed, routing, deadlock.
\end{IEEEkeywords}

\ifCLASSOPTIONpeerreview
\begin{center} \bfseries © 2020 IEEE. Personal use of this material is
permitted. Permission from IEEE must be obtained for all other uses,
in any current or future media, including reprinting/republishing
this material for advertising or promotional purposes, creating new
collective works, for resale or redistribution to servers or lists,
or reuse of any copyrighted component of this work in other works.
\end{center}
\fi

\IEEEpeerreviewmaketitle

\section{Introduction}

%
%
High performance computing (HPC) systems targeted for general purpose
applications require a large amount of computing resources. To achieve
this goal, interconnection networks play a key role as they need to
support efficient communication.

We can find many proposals for HPC systems based on an interconnection
network using a specific technology. Some of the most used technologies
are InfiniBand\cite{iba_2015} and Ethernet standard IEEE 802.3
introducing 40Gbps, 100Gbps, 200Gbps and 400Gbps data rates, which have
become the prevailing interconnection technologies for HPC systems. The
June 2019 list of top 500 supercomputer\cite{top500june2019} contains
HPC machines with over 2 million cores within the top 10
positions using the previous interconnection technologies.

Besides HPC systems, massively parallel computers (MPC) also require a
large amount of computing resources. However, MPCs are optimized for
specific applications (usually scientific computations). Many MPC
applications examples can be found in the literature, such as artificial
neural networks modeling\cite{nordstrom1992using} and quantum computer
simulations\cite{de2007massively}.

In these scenarios, topology changes due to application domain
changes becomes a critical aspect, specially in MPCs, regarding the
interconnection network. In addition, defective components may turn a
regular topology into an irregular one. Addressing these changes in
topology can lead to flawless operation and a more efficient utilization
of the interconnection network.

%
%
Furthermore, many-core processors are becoming increasingly popular as
VLSI circuits integration scale improves, allowing for billions of
transistors to be integrated within a die. We can find several
commercial products of these processors, such as the U.T. Austin
TRIPS\cite{gratz2007chip}, Intel Teraflops\cite{vangal200780}, Tilera
TILE64\cite{wentzlaff2007chip} and Intel's 72-core Knights Landing
co-processor\cite{sodani2016knights}. More recently, Cerebras launched a
400.000 Sparse Linear Algebra (SLA) cores chip using a 2D mesh on-chip
communication fabric\cite{fricker2019apparatus}.

One key challenge of many-core processors is the interconnection layer.
Bus and crossbar structures do not scale well as network size increases,
they may provide low bandwidth and high power consumption. To overcome
these limitations, packet switched networks-on-chip (NoCs) are used.
Nevertheless, to enable data transmission, these NoCs need additional
resources (e.g. buffers) and control logic (e.g. allocators) which may
become scarce resources as the system grows in complexity.

Although power consumption can be mitigated to some extent by many-core
architectures, transistor count will eventually raise the power
consumption challenge requiring novel low-power designs. These power
limitations may require transistors to be turned off (a.k.a. dark
silicon phenomenon\cite{esmaeilzadeh2011dark}) in order to keep hardware
within a given thermal design power envelope.

On the other hand, seeking new alternatives to large scale lithographic
transistors is attracting research interest. Nanoelectronics arise as a
feasible alternative, nevertheless they are prone to defects and
transient faults\cite{haselman2009future}. Fault-tolerance techniques
will be needed for this technology to be suitable for many-core
architectures.

%
%
The aforementioned challenges can be addressed using topology-agnostic
routing algorithms regardless of the environment. Thus,
topology-agnostic routing algorithms are presented as a suitable
solution to deal with changes in topology due to failure of components,
application domain changes and specific traffic needs. These algorithms
in combination with static or dynamic reconfiguration algorithms such
as OSR\cite{lysne2008efficient} and BLINC\cite{lee2014brisk} would
provide interconnection networks the ability to handle these situations
effectively.

In this paper we present Transparent Distributed Segment-based
Routing algorithm (TDSR), a topology-agnostic distributed routing
algorithm following the same working principle as the centralized
algorithm Segment-based Routing (SR) \cite{mejia2006segment}. The main
contribution of TDSR is to perform in a distributed environment without
the need for a centralized entity driving the entire process. TDSR
also provides a finer control on the segmentation process through the
assignment of link weights suitable for both on-chip and off-chip
networks. Using link weights TDSR brings the ability to adapt to
heterogeneous environments and a great variety of network requirements
brought by resource availability or application domain.

We analyze the performance in terms of execution time required for
the distributed algorithm to complete the segmentation process. In
particular, we show how different aspects such as the link weight
distribution and defective links rate have an impact on the algorithm
performance.

The rest of this paper is organized as follows. In Section
\ref{sec:background} we give a concise explanation of the necessary
concepts to understand the methodology. Next, in Section
\ref{sec:related-work} we provide a brief description of existing
topology-agnostic routing algorithms in the literature. Afterwards, in
Section \ref{sec:tdsr} we describe the TDSR algorithm. Finally, in
Sections \ref{sec:evaluation} and \ref{sec:conclusion} we evaluate and
analyze the different aspects driving TDSR performance and we draw some
conclusions as well as future work.

\section{Background} \label{sec:background}

In this section we cover the basic principles and concepts
regarding Minimum-weight Spanning Trees (MST) distributed
computation\cite{gallager1983distributed}, Lowest Common Ancestor (LCA)
identification and Segment-based Routing (SR)\cite{mejia2006segment}.
Our proposal will use these concepts as the building blocks to achieve
its goal (see Section \ref{sec:tdsr}).

\subsection{A Distributed Algorithm for Minimum-weight Spanning Trees}%
\label{sec:background:ghs}

Several algorithms for finding the Minimum-weight Spanning Tree (MST)
can be found in the literature. Classical approximations such as
Borůvska's\cite{nevsetvril2001otakar}, Prim's\cite{prim1957shortest},
and Kruskal's\cite{kruskal1956shortest} are considered well-suited for
many types of graphs $G = (V, E)$ achieving a time complexity of $O(|E|
\log{(|V|)})$\cite{chung1996parallel}. These algorithms, however, often
use elaborated data structures which prevent them from being used in a
distributed environment.

Finding a MST in a distributed fashion has been a research subject since
1977 with Spira's algorithm\cite{spira1977communication} followed by
Gallager's (GHS)\cite{gallager1983distributed} and later proposals based
on Gallager's algorithm\cite{awerbuch1987optimal, garay1998sublinear}.

Distributed algorithms are often characterized by its total
communication cost (a.k.a. communication complexity ($CC$)), defined
as the total amount of bits that participants of a communication
system need to exchange to perform a given task. Considering the set
of messages transmitted upon algorithm completion as $M = \{m_0,
m_1,...,m_{k-1}\}$, then:

\begin{displaymath}
CC = \sum\limits_{i=0}^{k-1}%
BitLength(m_{i})\times{}LinksTraversed(m_{i})
\end{displaymath}

For a connected undirected graph comprising $|V|$ vertices and
$|E|$ edges, GHS-based algorithms are optimal in terms of CC, with
an upper bound of $O(|E|+|V|log(|V|))$ messages. The original GHS
algorithm\cite{gallager1983distributed} requires unique finite weights
assigned to each edge. Then, messages transmitted by GHS contain at most
one edge weight plus $log_{2}(8|V|)$ bits. Time complexity upper bound
for the GHS algorithm is $O(|V|^2)$ in the general case. However, if all
vertices start computing the MST initially, this upper bound becomes
$O(|V|log(|V|))$.

GHS distributed algorithm assumes no central entity knowing the
properties of the graph. Instead, vertices initially know the weight of
their adjacent edges.  Nodes collaborate by exchanging messages over
adjoining links to construct the MST. GHS algorithm requires messages to
be transmitted independently in both directions at any given edge. In
addition, messages must arrive after an unpredictable but finite delay
lacking any error. Finally they must arrive in sequence. Out of order
message delivery shall not be supported by this algorithm.

In order to explain the idea behind GHS we must first introduce the
following definitions:

\begin{definition} Let $G$ be the edge-weighted undirected connected
graph that represents an interconnection network, denoted by $G = (V(G),
E(G))$, where $V(G)$ is the vertex set representing the nodes, and
$E(G)$ is the edge set representing the bidirectional physical links.
\end{definition}

\begin{definition} Each edge $e_i$ of $G$ can be expressed as a
$3$-tuple $(v_a, v_b, w_i)$ for $v_a, v_b \in V(G)$, where the unordered
pair of vertices $v_a$ and $v_b$ are the endpoints of $e_i$, and $w_i$
is an unique weight assigned to each edge $e_i$. \end{definition}

\begin{definition} Let $T$ be the minimum weight spanning tree (MST) of
$G$, denoted by $T = (V(T), E(T))$.  Hence, $V(T) \subseteq V(G)$ and
$E(T) \subseteq E(G)$. \end{definition}

\begin{definition} A fragment $F_j$ is defined as a subtree of $T$, that
is, for each $F_j$, $V(F_j) \subseteq V(T)$ and $E(F_j) \subseteq E(T)$.
Any two different fragments $F_j, F_k$ are vertex and edge disjoint,
such that, $V(F_j) \cap V(F_k) = \emptyset $ and $E(F_j) \cap E(F_k) =
\emptyset $. \end{definition}

\begin{definition} Let $F_j$ be a fragment. A fragment core $C_j$
is defined as a subtree of $F_j$ consisting of a single edge $e_i
\in E(F_j)$ and its endpoints. For a single vertex fragment $F_j$,
$C_j$~=~$F_j$. \end{definition}

Briefly, a \emph{fragment} of a MST is a subtree of the MST, i.e.
a connected subset of vertices and edges within the MST. An edge is
considered as an \emph{outgoing edge} of a fragment given that only
one of its adjacent vertices lies within the fragment. Using the previous
definitions, the following properties arise regarding MSTs:

\begin{property} Given a fragment of a MST, let $e$ be a minimum-weight
outgoing edge of the fragment. Then joining $e$ and its adjacent
non-fragment vertex to the fragment yields another fragment of a MST.
\end{property}

\begin{property} If all the edges of a connected graph have different
weights, then the MST is unique. \end{property}

A formal proof of these properties is provided by its authors in the
original GHS paper\cite{gallager1983distributed}.

The GHS algorithm differentiates three edge types, namely \emph{tree},
\emph{internal} and \emph{external} edges. \emph{Tree} edges, as the
name implies, belong to $T$.

\begin{definition} An internal edge $e'_i = (v_x, v_y, w_i)$ is an
edge of $G$ not belonging to $T$ whose vertices belong to the same
fragment. Hence, $e'_i \in E(G) \setminus E(T)$, $v_x \in V(F_j)$ and
$v_y \in V(F_j)$. \end{definition}

\begin{definition} An external edge $e''_i = (v_x, v_y, w_i)$ is an edge
of $G$ that connects two different fragments $F_j, F_k$.  Therefore,
$v_x \in V(F_j)$ and $v_y \in V(F_k)$ for $j \ne k$. \end{definition}

\begin{figure*}[!t]
\centering
\subfloat[Before merge.]{
    \includegraphics[width=2.5in]{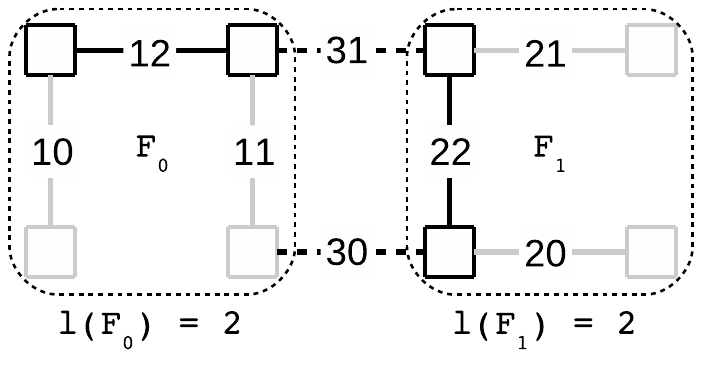}
    \label{fig:ghs-merge-example:a}
}
\hfil
\subfloat[After merge.]{
    \includegraphics[width=2.5in]{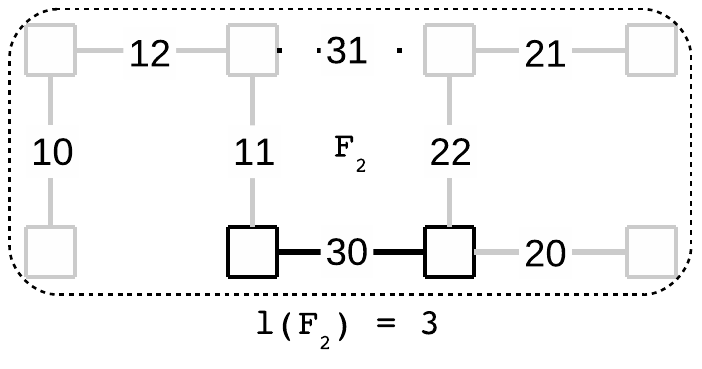}
    \label{fig:ghs-merge-example:b}
}
\hfil
\subfloat{
    \includegraphics[width=1.0in]{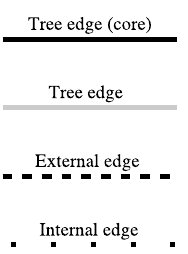}
}
\caption{Example of fragment merge. a) Fragments $F_0$ and $F_1$ have
equal level $l(F_0) = l(F_1) = 2$. b) A new fragment $F_2$ is created
and its core is moved towards the joint edge.}
\label{fig:ghs-merge-example}
\end{figure*}

During execution of the algorithm, \emph{external} edges may
become either \emph{internal} or \emph{tree} edges. Fig.
\ref{fig:ghs-merge-example} shows two external edges edges between two
different fragments, edges have weights $31$ and $30$ respectively.

In addition, each fragment $F_j$ has a \emph{level} associated, denoted
as $l(F_j)$. A single vertex fragment is defined to be at level $l(F_j)
= 0$.  Fragment level $l(F_j)$ is updated when two different fragments
$F_j$ and $F_k$ are joined together.

GHS algorithm starts with $N$ fragments made of single
vertices. Property 1 allows fragments to grow by joining other fragments
through their minimum weight outgoing edges, giving as a result a new
fragment of a MST. Moreover, Property 2 guarantees that the combined
fragment belongs to the same MST as both original fragments since
there is only one possible MST. Two cases arise when joining fragments
depending on their associated level:

\paragraph{Fragment merge}

Fig. \ref{fig:ghs-merge-example} shows an example of the distributed
algorithm for a given configuration of edge weights and fragments. Two
fragments $F_0$ and $F_1$ are already constructed. Each fragment core
is highlighted and both fragments have level $2$ associated. This
example shows how fragments at same level are merged. The outgoing
edge (dashed) for both fragments is edge weighted $30$ as Fig.
\ref{fig:ghs-merge-example:a} shows. Upon agreement on the outgoing
edge, $F_0$ and $F_1$ merge becoming the new fragment $F_2$ shown in
Fig. \ref{fig:ghs-merge-example:b}. Notice that $F_2$ core is located at
the joint edge and $l(F_2) = l(F_0)+1$ (or $l(F_2) = l(F_1)+1$).

\paragraph{Fragment absorption}

\begin{figure*}[!t]
\centering
\subfloat[Before absorption.]{
    \includegraphics[width=2.5in]{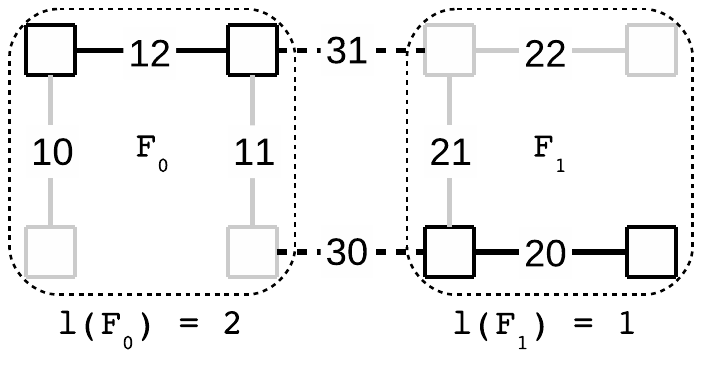}
    \label{fig:ghs-absorption-example:a}
}
\hfil
\subfloat[After absorption.]{
    \includegraphics[width=2.5in]{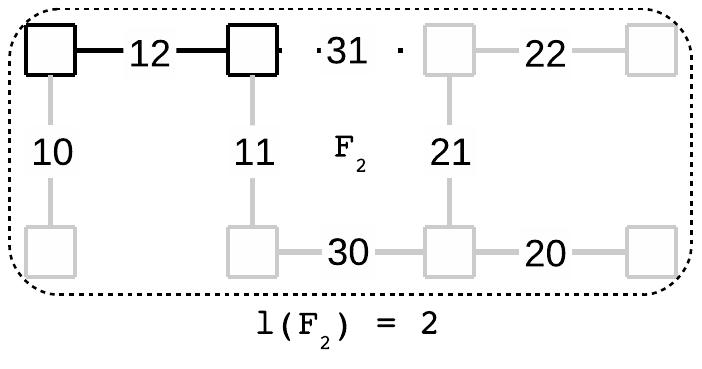}
    \label{fig:ghs-absorption-example:b}
}
\hfil
\subfloat{
    \includegraphics[width=1.0in]{gfx/ghs-example-legend-eps-converted-to.pdf}
}
\caption{Example of fragment absorption. a) Fragment $F_1$ has lower
level than $F_0$. b) $F_1$ is absorbed by $F_0$ creating a new fragment
$F_2$ which inherits $F_0$ core and also its level.}
\label{fig:ghs-absorption-example}
\end{figure*}

Fig. \ref{fig:ghs-absorption-example:a} shows a similar example with
two fragments $F_0$ and $F_1$ already set. Again, fragment cores are
highlighted in black but this time fragment levels are different,
$l(F_0) = 2$ while $l(F_1)~=~1$. Once $F_1$ selects its outgoing edge
weighted $30$, $F_1$ is absorbed by $F_0$ becoming $F_2$ as Fig.
\ref{fig:ghs-absorption-example:b} shows. Absorption occurs because
$l(F_0) > l(F_1)$. Notice that fragment absorption does not require the
greater level fragment to agree upon the outgoing edge selected by the
lower level fragment. This prevents lower level fragments to wait for
greater level fragments to select the same outgoing edge, decreasing the
time required by the distributed algorithm. In this case, $F_2$ inherits
$F_0$ core and also its level.

An interesting property of the algorithm is that MST construction is
driven by the edge's weights distribution. Therefore, different weights
distributions may achieve different MST configurations.

\subsection{Ancestor Identification Labeling Scheme}%
\label{sec:background:labelling}

Given the previously computed MST ($T$), we traverse it in a pre-order
fashion assigning an integer $a_i$ to each traversed vertex $v_i$
in a strictly-increasing monotonic order. Additionally, each $v_i$
is assigned an integer $b_i$ as the maximum $a_j$ among all $v_i$
successors. Notice that $a_i = b_i$ if vertex $v_i$ is leaf in $T$.

\begin{definition}
For each vertex $v_i$ of T, there exists a different label $t(v_i) = [a,
b]$ which is a closed integer interval with lower and upper bounds $a$
and $b$ respectively.
\end{definition}

By construction, for any given pair of distinct vertices $v_i$ and
$v_j$, this labeling scheme satisfies the following property.

\begin{property}
$t(v_j) \subset t(v_i)$ iif $v_i$ is an ancestor of $v_j$. In other
words $t(v_j)$ is a subinterval of $t(v_i)$.\label{pro:ancestor}
\end{property}

Using Property \ref{pro:ancestor}, the Common Ancestor (CA) definition
can be expressed as:

\begin{definition}
Given $v_i, v_j, v_k$ of T, $v_i$ is a Common Ancestor (CA) of vertices
$v_j$ and $v_k$ iif $t(v_j) \subset t(v_i)$ and $t(v_k) \subset t(v_i)$.
\label{def:ca}
\end{definition}

\begin{figure}[!t]
    \centering
    \includegraphics[width=1.5in]{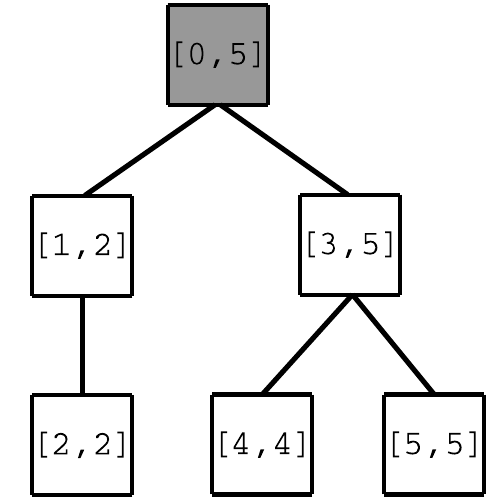}
    \caption{
        Label computation example.
    }
    \label{fig:labelling-example}
\end{figure}

Fig. \ref{fig:labelling-example} shows an example of label computation
in a rooted tree. Intervals lower bound are computed in a pre-order
fashion from the tree's root. Then, upper bounds are computed from leaf
vertices towards the root. A similar approach is followed by Santoro et
al.\cite{santoro1985labelling} to provide an implicit routing for graphs
representing arbitrary topologies.

\subsection{Segment-based Routing}\label{sec:background:sr}

Segment-based Routing (SR) \cite{mejia2006segment} is a
topology-agnostic routing algorithm aimed to provide a reasonable
path quality and fault-tolerance while keeping complexity low. SR is
considered a rule-driven routing algorithm \cite{flich2011survey} whose
main features and required resources can be summarized as follows:

\begin{enumerate}
    \item It does not guarantee shortest path computation by design.
    \item Virtual channels are not required.
    \item Deadlock freedom enforcement and path selection stages do not
rely on spanning tree computation.
\end{enumerate}

SR working principle is the partitioning of a topology into
subnets. Subnets in turn are made of disjoint segments where routing
restrictions are placed locally within each segment to break cycles,
thus guaranteeing deadlock freedom and connectivity within each
subnet. During the partitioning process, network links are visited at
most once to avoid the procedure to reach a deadlocked state.

A segment is defined as a list of interconnected switches and
links. Three different segment types can arise from partitioning:

\begin{itemize}
    \item Starting segment: It starts and ends at the same switch (i.e.
it makes a cycle). SR begins the partitioning process by building
this segment. Thus, we have to choose a switch (i.e. starting switch)
and find a cycle that includes it. The cycle itself is considered the
starting segment. There is one starting segment per subnet, which may
connect to a different subnet through a \emph{bridge} link.
    \item Regular segment: These segments start on a link followed by
one or more switches and links, and they end on a link.
    \item Unitary segment: They are made by a single link.
\end{itemize}

Guaranteeing deadlock-freedom and connectivity while preserving segments
independent from each other, requires the fulfillment of the following
construction rules\cite{mejia2006segment}:

\begin{myrule}
    Nodes and links can be included in no more than one segment.
    \label{rule:sr-1}
\end{myrule}

\begin{myrule}
    New regular or unitary segments must be connected to previous
constructed segments. Starting segments do not follow this rule.
    \label{rule:sr-2}
\end{myrule}

\begin{figure}[!t]
    \centering
    \includegraphics[width=2.5in]{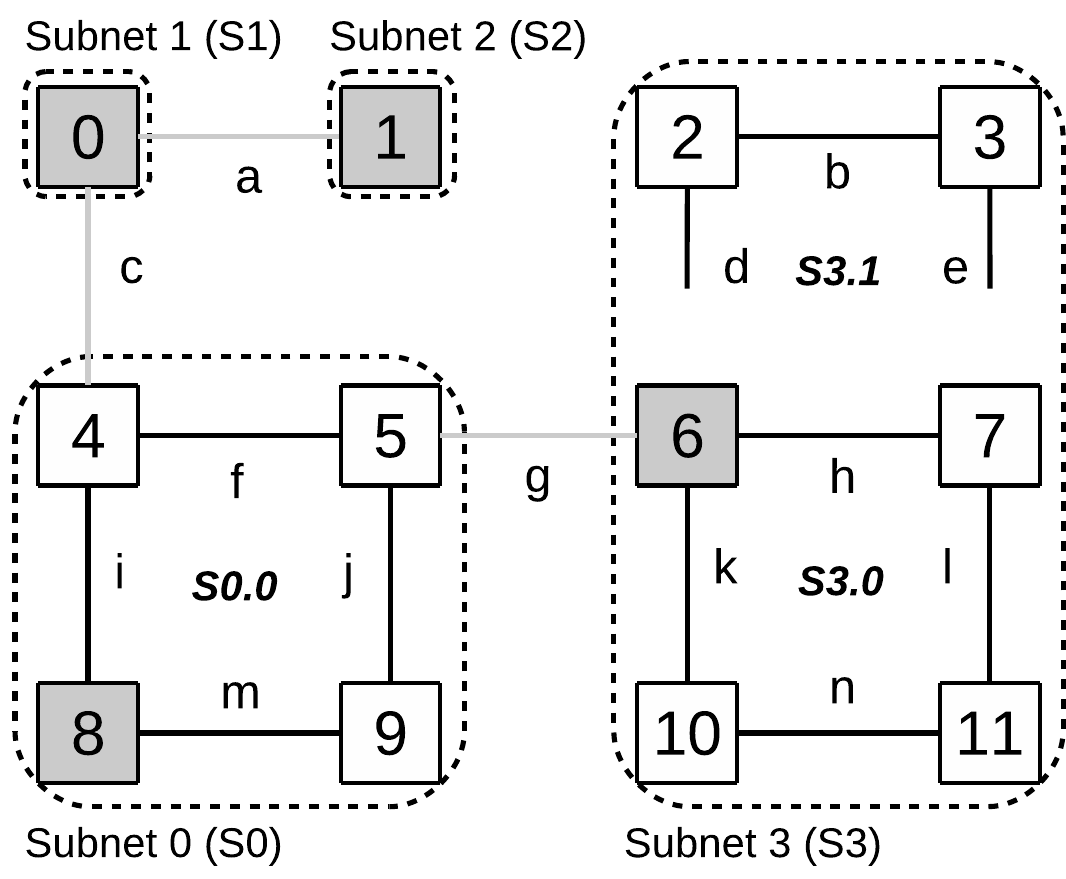}
    \caption{SR partitioning example with four subnets (dashed areas)
    and three segments S0.0, S3.0 and S3.1. Starting switches within
    each subnet are filled in gray. Bridge links, namely $a$, $c$ and
    $g$ are also showed.}
    \label{fig:sr-example}
\end{figure}

Fig. \ref{fig:sr-example} shows an example of computed subnets and
segments in a 4x3 Mesh with some links missing. In this case, we have
four subnets (dashed areas) and three regular segments $S0.0 = (\{8,
4, 5, 9\}, \{i, f, j, m\})$ , $S3.0 = (\{6, 7, 11, 10\}, \{h, l, n,
k\})$ and $S3.1 = (\{2, 3\}, \{d, b, e\})$. Grayed links joining two
subnetworks are not included in any segment, they are considered
\emph{bridge} links. 

In Fig. \ref{fig:sr-example} example, switch 8 was
chosen as the starting switch. Then, construction of subnet 0 (S0)
begins by building the starting segment within the subnet, that is,
finding a cycle from switch 8, yielding segment $S0.0$.

Once the starting segment is found, the algorithm selects among the
available outgoing links sourcing previously constructed segments. In
this case, only $c$ and $g$ links meet this condition, we choose $c$.

From link $c$, we look for paths ending at any switch within a previous
constructed segment. With lack of success, $c$ link is considered a
\emph{bridge} and a new subnet S1 is started, becoming switch 0 the
starting switch for the newly created subnet S1.

As each link can be visited only once, to find a cycle from switch 0,
we are forced to choose link $a$, where no cycle is found, letting $a$
being a \emph{bridge} link and a new subnet S2 is started, making switch
1 its starting switch.

In a similar fashion, link $g$ is marked as \emph{bridge}, triggering
the construction of a new subnet S3 with switch 6 as the starting
switch. This time, a cycle is found through link $h$ and the starting
segment for S3 is built ($S3.0$).  Then, we look for available outgoing
links sourcing previously constructed segments within the subnet, i.e.
$\{d, e\}$ links. By choosing $d$, a path is found to end at a switch
already within a previously built segment (switch 7 through $e$), this
path becomes the last segment $S3.1$. We provide a detailed SR
specification in Appendix \ref{sec:sr}.

\section{Related Work}\label{sec:related-work}

Topology-agnostic routing algorithms can be classified in two major
categories regarding the computation of the final set of paths
between end-points\cite{flich2011survey}. \emph{Path-driven} routings
achieve this task by selecting among the initial set of paths
meeting certain criteria (path length, traffic balance, etc.). The
second step focuses on guaranteeing deadlock-freedom, often through
the use of virtual-channels to break cyclic path dependencies.
TOR\cite{sancho2002effective}, LASH\cite{skeie2002layered} and
LASH-TOR\cite{skeie2004lash} are some of the routing algorithms falling
into the \emph{path-driven} category.

On the other hand, in \emph{rule-driven} routings the first step is
to achieve deadlock-free routes by the use of rules to break cyclic
dependencies, once deadlock freedom is ensured, they must select
paths between each source-destination pair to obtain a deterministic
routing based upon some criteria (random, round-robin, minimal path,
etc.). UD\cite{schroeder1991autonet}, FX\cite{sancho2000flexible},
LTURN\cite{koibuchi2001turn}, SR\cite{mejia2006segment}, \emph{Tree-turn
Routing}\cite{zhou2012tree} and DiSR\cite{catania2014distributed} fall
into the \emph{rule-driven} routing algorithms category.

Due to the nature of the functional steps performed by
\emph{rule-driven} algorithms, they are likely to be implemented in a
distributed fashion using only local information at each communication
node, i.e. some aspects such as the complete topology and status of
the network are not available. \emph{Path-driven} algorithms, however,
require the complete set of paths between each source-destination pair
to be able to select among them so as to break cyclic dependencies.

Distributed computation of topology-agnostic routing algorithms enables
a more scalable, resilient and flexible routing algorithm. DiSR
for instance, is an attempt to achieve the same goals as the SR
algorithm but it does not rely on a topology graph nor a centralized
computation. Nevertheless, DiSR focuses on fast computation of
routing restriction rules rather than guaranteeing connectivity among
switches. Thus, leaving unconfigured switches even when computation
is performed on a regular topology due to its non-deterministic
nature. Traffic routed through unconfigured switches may result in
deadlock situations due to the lack of routing restrictions. As a
consequence, the system is forced to avoid routing traffic through
unconfigured switches.

So far we have discussed \textit{deadlock avoidance} routing algorithms.
Nevertheless, \textit{deadlock recovery} topology-agnostic mechanisms
have also been proposed\cite{anjan1995efficient, ramrakhyani2017static,
ramrakhyani2018synchronized}. These algorithms aim to detect deadlock
situations at runtime and recover from it. The reason behind these
proposals is that deadlocks are not likely to occur very often. However,
tracking network cyclic dependencies require complex logic as well as
the mechanism to recover from it. Besides, deadlock detection often
relies on counter thresholds which must be properly configured in order
to reduce the amount of false positives.

The TDSR approach inherits most of the features of the SR algorithm as
well as \emph{deadlock avoidance} \emph{rule-driven} algorithms. In
addition, TDSR deterministic computation is performed in a completely
distributed fashion guaranteeing optimal coverage of switches into
segments also in presence of connected components in the topology.

\section{Transparent Distributed Segment-Based Routing}\label{sec:tdsr}

In this section we present our proposal, namely Transparent Distributed
Segment-Based Routing (TDSR). We give a brief overview of the main
goal and features of the proposed distributed algorithm. Then, we
provide a detailed explanation of the functional stages performed by
TDSR: \emph{MST construction}, \emph{labeling} and \emph{segment
construction}.

The main goal of TDSR is devised from the SR\cite{mejia2006segment}
original proposal i.e., partitioning of a topology into subnets which,
in turn, are further divided in segments.

As we explained in Section \ref{sec:background}, SR requires the full
set of switches and links aimed to be partitioned, then, it starts
searching for segments using these two sets. This general approach
is not suitable to be carried out in a distributed environment where
only local information is available at each communication node (a.k.a.
switch). TDSR proposal is a distributed approach to achieve the same
objective.

Some SR features are also provided by TDSR:

\begin{itemize}
    \item It does not guarantee minimal routing by design.
    \item Virtual channels are not required.
\end{itemize}

We can split TDSR in three functional stages:

\begin{enumerate}
    \item \emph{MST construction} in order to enforce deadlock freedom.
    \item \emph{Switch labeling} to enable Lowest Common Ancestor computation
(LCA).
    \item \emph{Segment construction} using the information provided by
previous stages.
\end{enumerate}

\subsection{Minimum Spanning Tree Construction Stage}%
\label{sec:mst-stage}

TDSR takes advantage of the structure provided by MSTs to identify
network links introducing cycles in the network graph (i.e. links not
included in the MST). Thus TDSR considers two different types of links,
those included in the computed MST (a.k.a. \emph{tree} links), and those
links not included in the MST (a.k.a. \emph{internal} links). There is
also a third link type which arises during the MST construction, called
\emph{external} link.  This third type of link connects switches within
different fragments of the MST which have yet to be joined (see Section
\ref{sec:background:ghs}).

We assume that each network link has a different unique weight.
Nevertheless, if the previous assumption is not fulfilled, unique link
weights can be computed using the identifiers of both switches connected
through each link\footnote{Assuming that switch identifiers are unique
within the network.}. Then, TDSR uses the GHS distributed algorithm
explained in Section \ref{sec:background:ghs} to construct the MST.

\begin{figure*}[!t]
    \centering
    \includegraphics[width=7in]{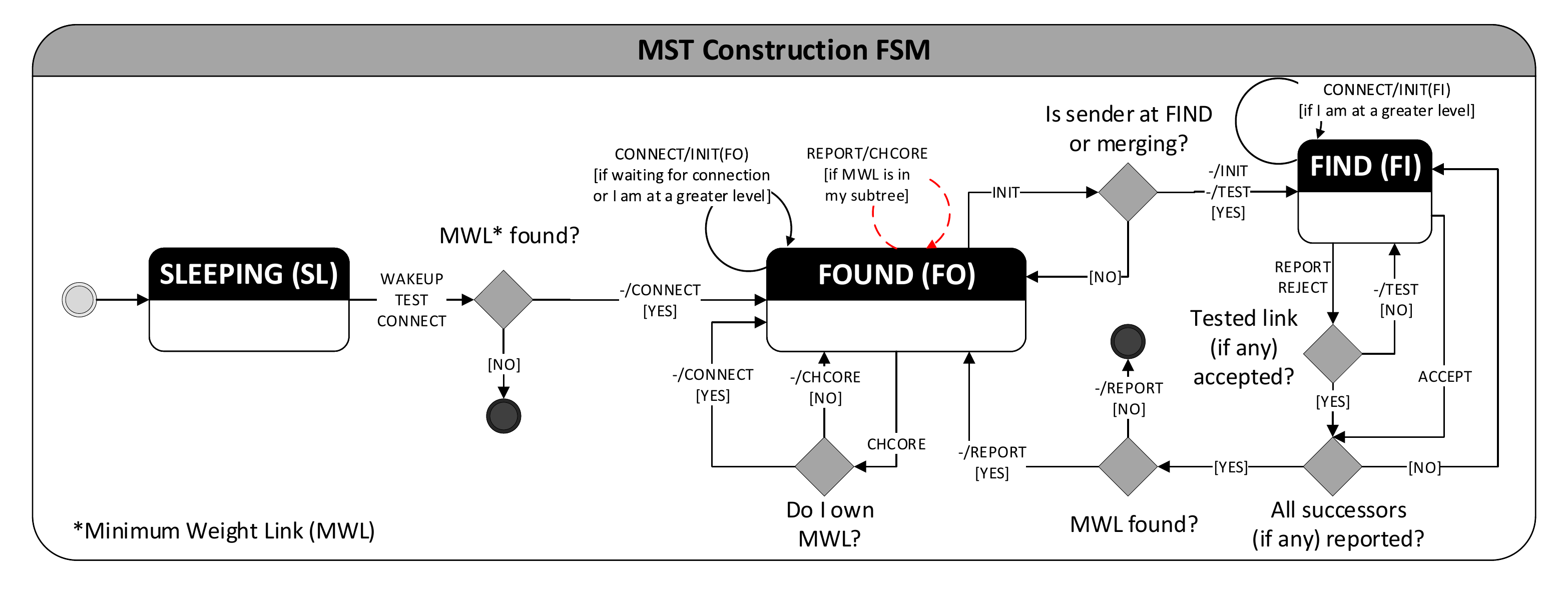}
    \caption{MST Construction FSM diagram.}
    \label{fig:ghs-fsm-diagram}
\end{figure*}

Fig. \ref{fig:ghs-fsm-diagram} shows the Finite-State Machine (FSM)
diagram for the MST construction stage. This automata is run by every
switch and it consists of three states, namely \emph{SLEEPING} (SL),
\emph{FOUND} (FO) and \emph{FIND} (FI). Stage transitions are triggered
upon arrival of messages, actions performed within state transitions may
involve sending new messages to adjacent switches.

\begin{figure*}[!t]
\centering
    \subfloat[]{
        \includegraphics[width=1.30in]{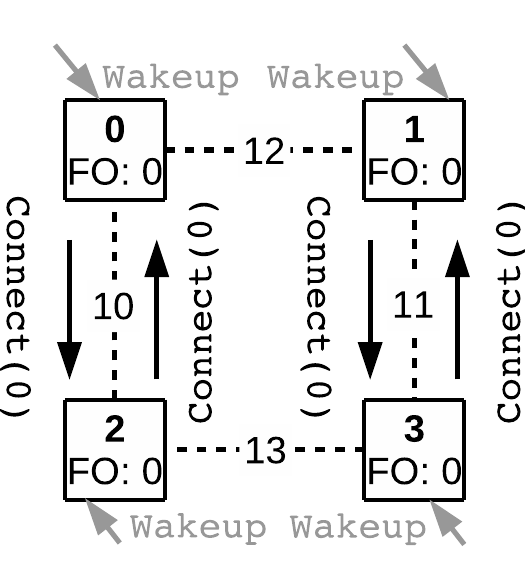}
    \label{fig:mstree-stage-example:a}}
\hfil
    \subfloat[]{
        \includegraphics[width=1.30in]{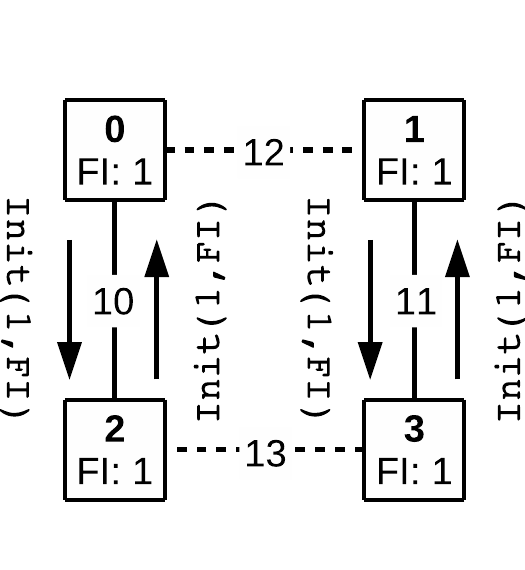}
    \label{fig:mstree-stage-example:b}}
\hfil
    \subfloat[]{
        \includegraphics[width=1.30in]{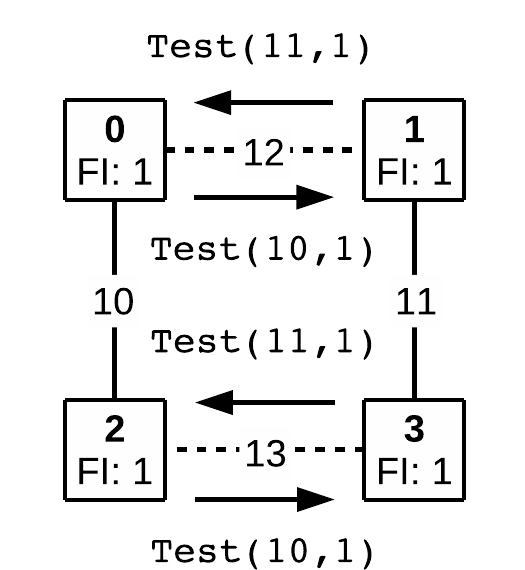}
    \label{fig:mstree-stage-example:c}}
\hfil
    \subfloat[]{
        \includegraphics[width=1.30in]{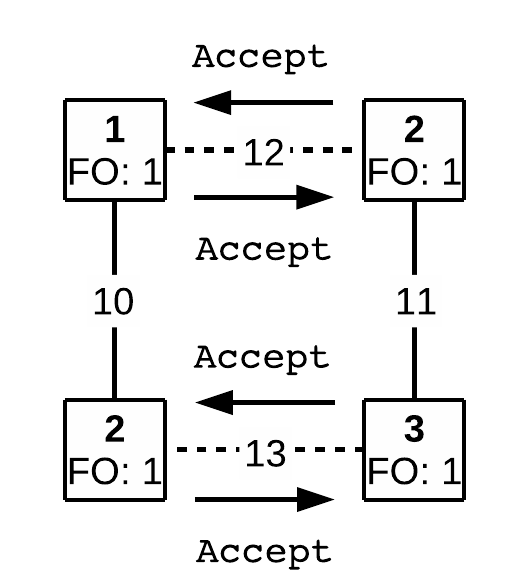}
    \label{fig:mstree-stage-example:d}}
\hfil
    \subfloat[]{
        \includegraphics[width=1.30in]{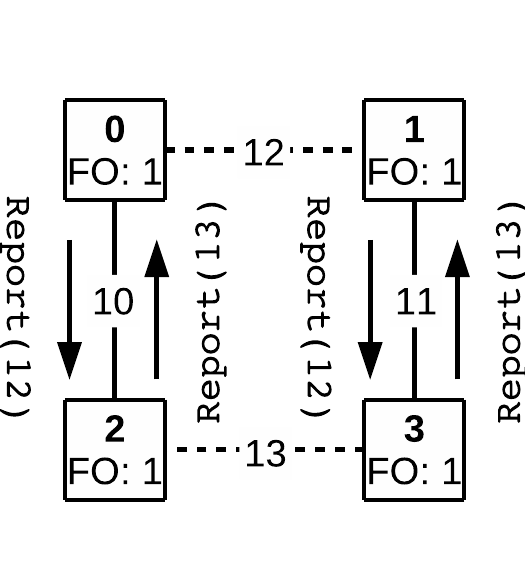}
    \label{fig:mstree-stage-example:e}}
\hfil
    \subfloat[]{
        \includegraphics[width=1.30in]{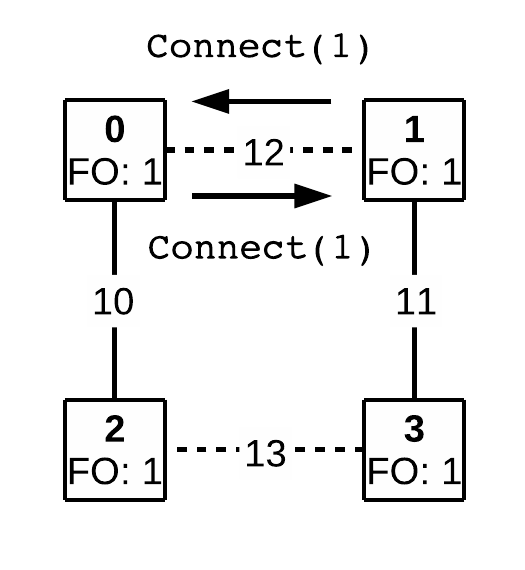}
    \label{fig:mstree-stage-example:f}}
\hfil
    \subfloat[]{
        \includegraphics[width=1.30in]{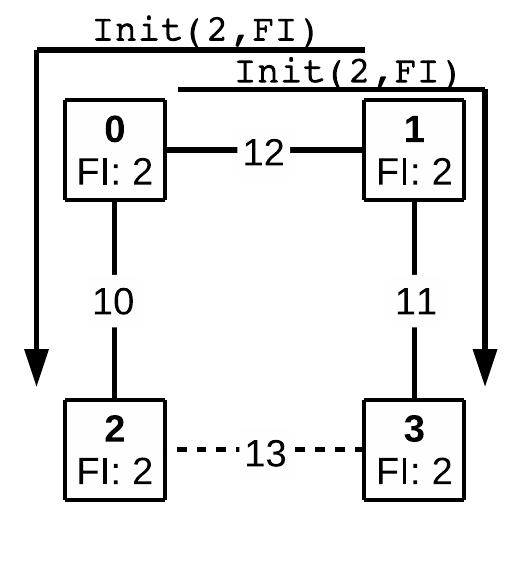}
    \label{fig:mstree-stage-example:g}}
\hfil
    \subfloat[]{
        \includegraphics[width=1.30in]{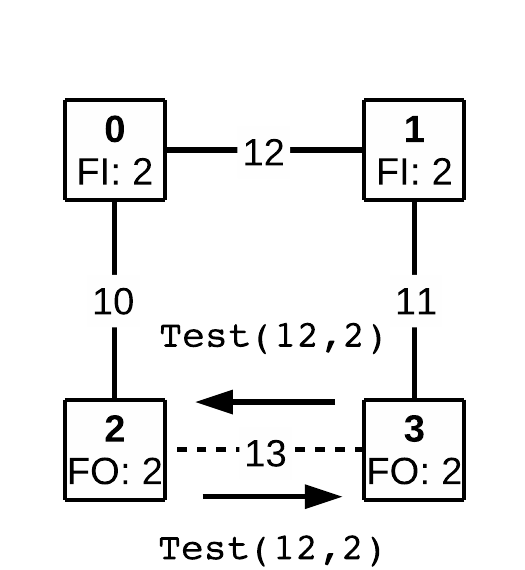}
    \label{fig:mstree-stage-example:h}}
\hfil
    \subfloat[]{
        \includegraphics[width=1.30in]{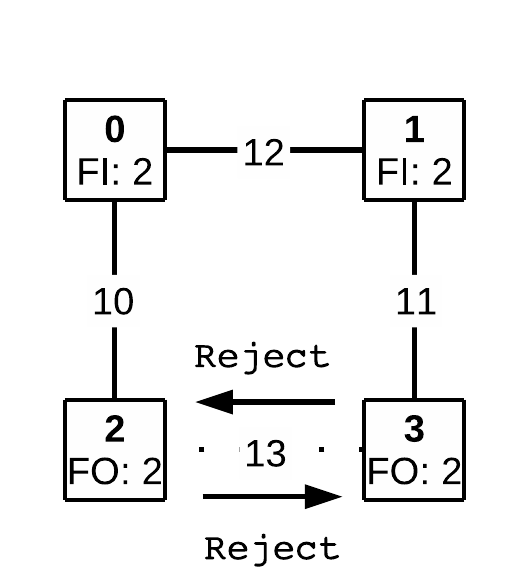}
    \label{fig:mstree-stage-example:i}}
\hfil
    \subfloat[]{
        \includegraphics[width=1.30in]{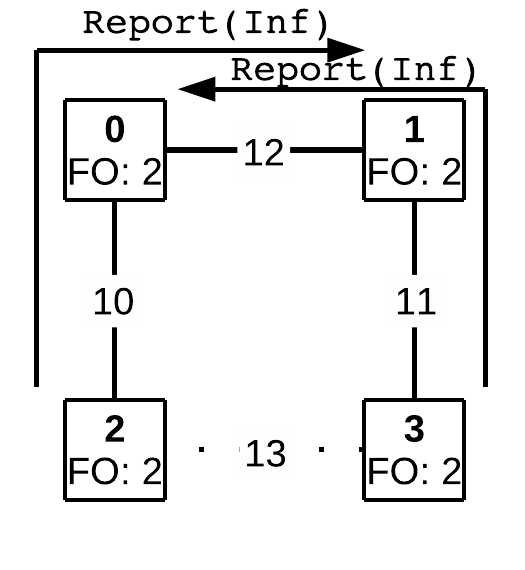}
    \label{fig:mstree-stage-example:j}}
\caption{MST Construction example. a), b) Initial level 0 fragments
merge. c), d), e) Level 1 fragments search for outgoing edges. f, g)
Level 1 fragments merge. h), i), j) Level 2 fragment tags edge 13 as
\emph{Internal} and finishes.}
\label{fig:mstree-stage-example}
\end{figure*}

There are eight types of messages involved in the MST construction. Each
message type is used at different steps of the procedure. In order
to understand each step and the involved messages, we show a small
example comprising four switches in a mesh arrangement (see Fig.
\ref{fig:mstree-stage-example}).

Initially all switches are in the SL state and they form a different
fragment by themselves of level $0$. All network links are considered
\emph{External}. At any given switch, upon arrival of either a
\emph{WAKEUP}, \emph{TEST} or \emph{CONNECT} message, it starts
searching for its minimum weight adjacent link. For simplicity we assume
that all switches in SL state receive a \emph{WAKEUP} message to trigger
the search for their minimum weight adjacent link.

If no link is found at this point means that the switch is isolated
from the network. On the other hand, if a Minimum Weight Link (MWL)
is found, the switch sends a CONNECT message containing its level
($0$) through the selected link to join the remote fragment (Fig.
\ref{fig:mstree-stage-example:a}). A \emph{WAKEUP} message is
sent through the remaining links (if any) and the switch moves to
state FO.  Hence, the transition from FSM state SL to FO in Fig.
\ref{fig:ghs-fsm-diagram} has been accomplished because the MWL has been
found within each level 0 fragment.

Each switch eventually receives its corresponding CONNECT message. Then,
they answer with an INIT message to merge with the remote fragment
as Fig. \ref{fig:mstree-stage-example:b} shows. At this point, two
fragments with switches $\{0, 2\}$ and $\{1, 3\}$ are derived from the
merge operation. The INIT message received at each switch causes the
transition from state FO to FI. It also changes the link type from
\emph{external} to \emph{tree} (solid links in the figure) considering
it part of the MST.

When transitioning to state FI, switches must find their lower
weight adjacent link locally. This is accomplished by sending
TEST messages through the lowest weight adjacent link as Fig.
\ref{fig:mstree-stage-example:c} shows. TEST messages contain the sender
fragment identifier and level. The fragment identifier is the link
weight of the fragment core. Hence, fragments $\{0, 2\}$ and $\{1, 3\}$
identifiers are $10$ and $11$ respectively.  For level 0 fragments, no
identifiers are considered.

Upon the arrival of a TEST message through an \emph{external} link,
receiving switches may reply either an ACCEPT or REJECT message if the
following conditions are met:

\begin{enumerate}
    \item If fragment identifiers of sender and receiver are equal, a
REJECT message is sent back.
    \item If fragment identifiers of sender and receiver are different,
and sender fragment level is lower or equal than receiver fragment
level, an ACCEPT message is sent back.
    \item If conditions 1 and 2 are not met, fragment identifiers of
sender and receiver must be different and sender fragment has greater
level than receiver fragment. In this case, the receiver delays the
reply until either condition 1 or 2 is met. Hence, the sender will stall
at state FI preventing the greater level fragment to proceed.
\end{enumerate}

Our example has both fragments $\{0, 2\}$ and $\{1, 3\}$ at the same
level ($1$) with different identifiers. Therefore, upon arrival
of TEST messages, switches send back an ACCEPT message as Fig.
\ref{fig:mstree-stage-example:d} shows. These ACCEPT messages trigger
the transition from state FI to state FO which result in REPORT messages
being sent towards core switches. REPORT messages contain the MWL found
within a branch (see Fig. \ref{fig:mstree-stage-example:e}). Notice that
core switches are not considered successors between themselves. Thus,
condition \emph{All successors (if any) reported?} within the transition
from state FI to state FO is met (see Fig. \ref{fig:ghs-fsm-diagram}).

MWL must be located at only one side of the fragment core, in
this case, the MWL selected by both fragments is link weighted
$12$. Downwards search through neither fragment is needed to locate
the MWL because it is located at one of the core switches within each
fragment. Otherwise, a CHCORE message would be needed to locate the
MWL adjacent switch\footnote{If the MWL is not connected to a core
switch, CHCORE messages would be sent downwards the fragment branch to
accurately locate the MWL adjacent switch within that branch (dashed
loop at FO state shown in Fig. \ref{fig:ghs-fsm-diagram}).}. Upon
reception of a CHCORE message at the MWL adjacent switch (not
needed if MWL adjacent switch is within the fragment core), it
sends a CONNECT message to merge with the remote fragment (see Fig.
\ref{fig:mstree-stage-example:f}).

Upon arrival of CONNECT messages, Fig. \ref{fig:mstree-stage-example:g}
shows INIT messages exchanged between remote fragments to create the
new fragment resulting from the merge of fragments $\{0, 2\}$ and $\{1,
3\}$. The new fragment $\{0, 1, 2, 3\}$ has level 2 with edge $12$ as
its core. Upon reception of the INIT message, switch move to the FI state
to begin searching for a new MWL sending the corresponding TEST messages
(see Fig. \ref{fig:mstree-stage-example:h}).

Tested link $13$ has both adjacent switches at the same
fragment. Therefore, it is considered an \emph{internal} link by means
of REJECT messages (see Fig. \ref{fig:mstree-stage-example:i}).

Finally, REPORT messages are sent back to the core notifying that no MWL
has been found (see Fig. \ref{fig:mstree-stage-example:j}). The MST
construction stage finishes upon arrival of REPORT messages with no
MWL found at switches through all its downward adjacent edges in the tree
as in Fig. \ref{fig:mstree-stage-example:j}. Once core switches receive
these REPORT messages through all its adjacent edges (including its
sibling core switch), we can safely assume that the final MST has been
constructed\cite{gallager1983distributed} (i.e. a single fragment
remains).

The MST construction example shown in Fig.
\ref{fig:mstree-stage-example} puts on display one of the properties
mentioned in Section \ref{sec:background:ghs}. Edge weights drive the
construction of the MST by definition. As a consequence, different
MST can be constructed upon different edge weight configurations.

\begin{figure*}[!t]
\centering
\subfloat[Horizontal distribution]{
    \includegraphics[width=0.30\textwidth]{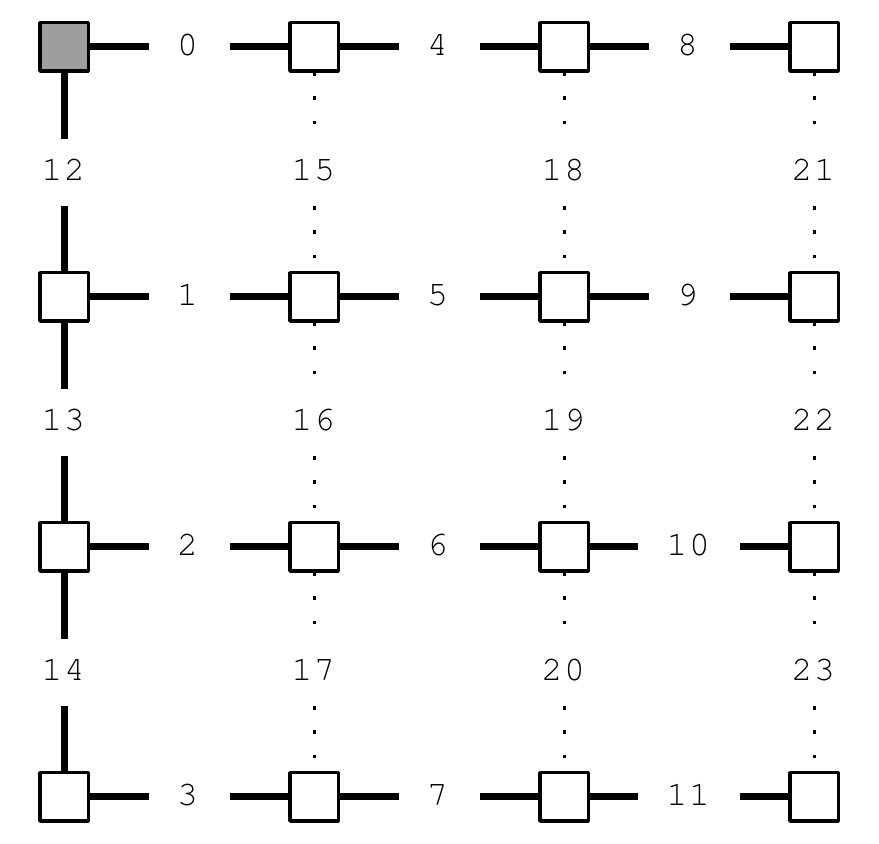}
    \label{fig:4x4-tree:hfirst}
}
\hfil
\subfloat[Center distribution]{
    \includegraphics[width=0.30\textwidth]{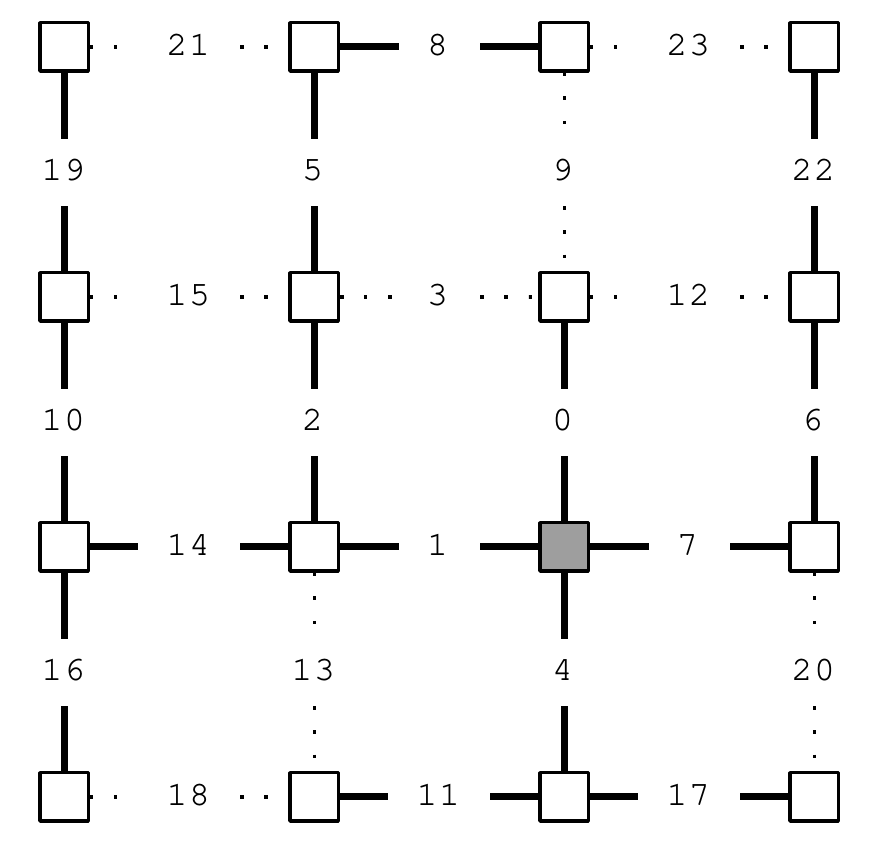}
    \label{fig:4x4-tree:center}
}
\hfil
\subfloat[Zigzag distribution]{
    \includegraphics[width=0.30\textwidth]{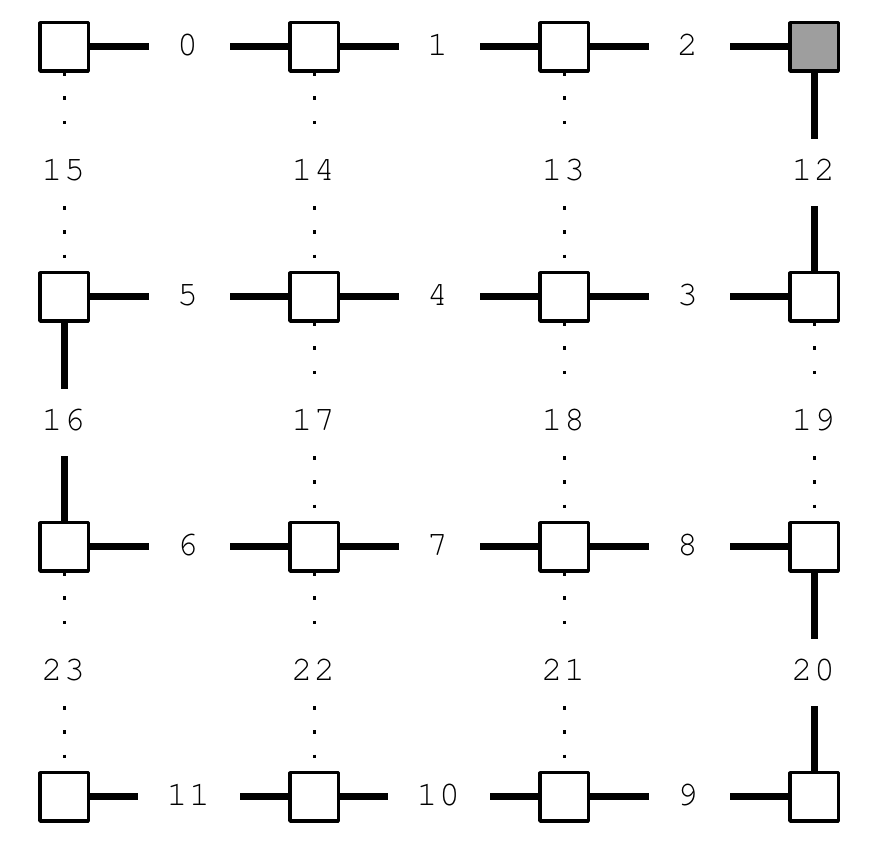}
    \label{fig:4x4-tree:zigzag}
}
\caption{Tree configurations derived from different edge's weight
distributions.}
\label{fig:ghs-example}
\end{figure*}

For instance, Fig. \ref{fig:ghs-example} shows different MST derived
from three different link weight configurations in a $4\times4$
mesh (grayed switches are one of the core switches chosen as tree
root located at the MST core\footnote{Note that once the MST is
found, only one fragment (the whole tree) remains.}). In Fig.
\ref{fig:4x4-tree:hfirst}, horizontal edges have lower weight than
vertical edges. Also, leftmost edges have a lower weight associated.

On the other hand, Fig. \ref{fig:4x4-tree:center} shows an edge weight
distribution from the mesh center towards the boundary. Finally, Fig.
\ref{fig:4x4-tree:zigzag} shows the MST constructed from an edge weight
configuration in a \emph{zigzag} fashion.

\subsection{Labeling Stage}

As soon as switches finish the MST Construction Stage (see
\ref{sec:mst-stage}), they proceed with the labeling stage. The
labeling stage objective is to provide each switch with the necessary
information to know whether it is ancestor of any given switch or
not. To achieve this, a closed integer interval is associated
to each switch, regarded as the switch's label. A more elaborate
explanation of the labeling scheme used is provided in Section
\ref{sec:background:labelling}.

\begin{figure*}[!t]
    \centering
    \includegraphics[width=7in]{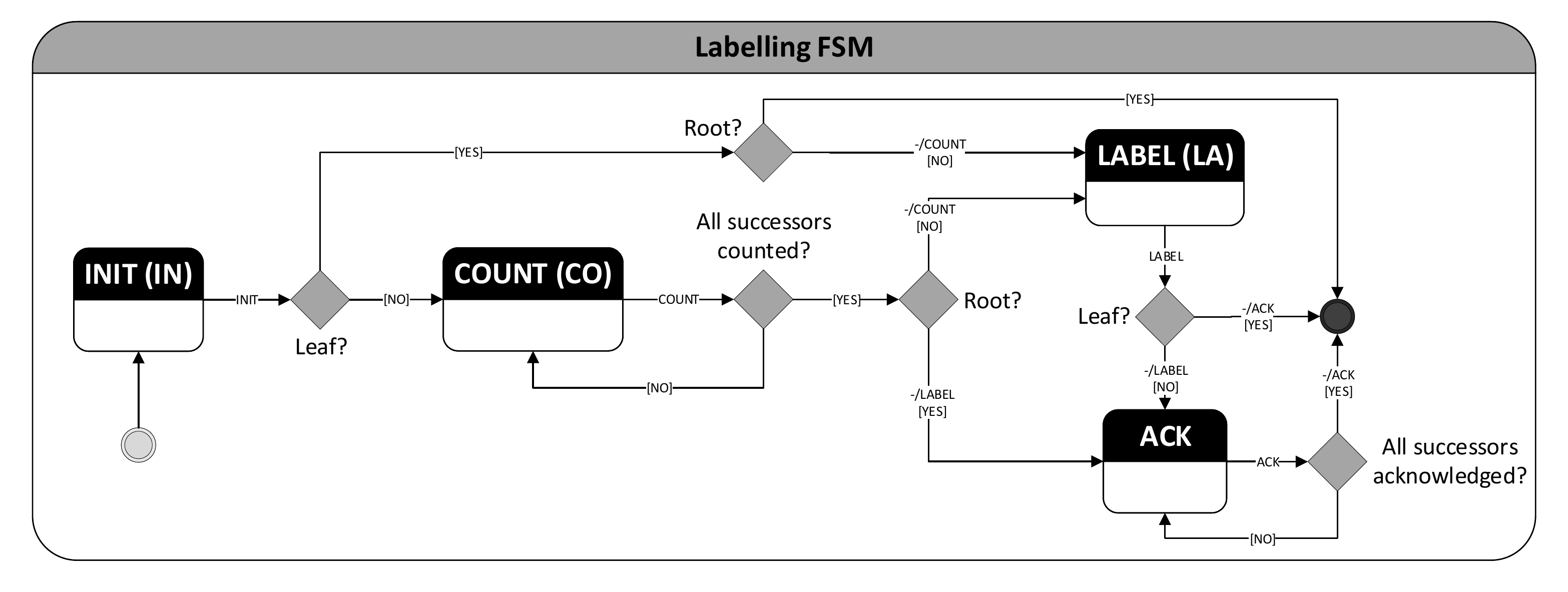}
    \caption{Labeling FSM diagram.}
    \label{fig:labelling-fsm-diagram}
\end{figure*}

The finite state machine diagram of this stage's implementation is provided in
Fig. \ref{fig:labelling-fsm-diagram}. It consists of four states: \emph{INIT}
(IN), \emph{COUNT} (CO), \emph{LABEL} (LA) and \emph{ACK}, each with its
corresponding message type named after it.

To compute interval lower bounds, each switch must account for the
successors in the MST (i.e. using \emph{tree} links) by means of COUNT
messages starting at the leafs, towards the root. Then, an integer
value representing the next hop lower bound is sent per downward link
starting at the root switch (LABEL messages). For the first downward
link considered, the next hop lower bound transmitted must be equal to
the switch's lower bound increased by one.

Lower bounds transmitted through the following downward links are
increased by the amount of successors reachable through previously
transmitted downward links. This is similar to depth-first search (DFS)
traversal algorithm.

Finally, upper bounds are computed at each switch by selecting the
maximum lower bound among its successors upon reception of ACK messages
from all its successors. Fig. \ref{fig:labelling-example} shows an
example with switch labels already computed.

Nodes are visited three times: upwards to propagate the amount of
successors towards the tree's root, then, downwards to compute each
switch's label. Finally, ACK messages are transmitted upwards to compute
intervals upper bound. Therefore, time complexity upper bound for this
stage is $O(3n)$.

\subsection{Segment Construction Stage}\label{sec:segmentation-stage}

The segment construction stage objective is to build segments using
the \emph{internal} links (i.e. links not belonging to the MST) arisen
from the MST construction stage. The segment build process follows SR
construction Rules \ref{rule:sr-1} and \ref{rule:sr-2} discussed in
Section \ref{sec:background:sr} to guarantee deadlock-freedom while
preserving connectivity. Segments are identified by the weight of the
\emph{internal} link that triggered the segment construction.

\emph{Internal} links are considered suitable to build a new segment
if both endpoints' \emph{Lowest Common Ancestor} (LCA) lies within
existing segments. Link suitability can be checked by ensuring that
switches connected to the constructed segments area (a.k.a. constructed
area) but not yet included are not \emph{Common Ancestors} (CAs) of both
link endpoints according to Definition \ref{def:ca}. This ensures that
the link endpoints LCA lies within the constructed area. Therefore, the
new segment will be connected to previous constructed segments and Rule
\ref{rule:sr-2} is satisfied.

\begin{figure}[!t]
\centering
\includegraphics[width=2.5in]{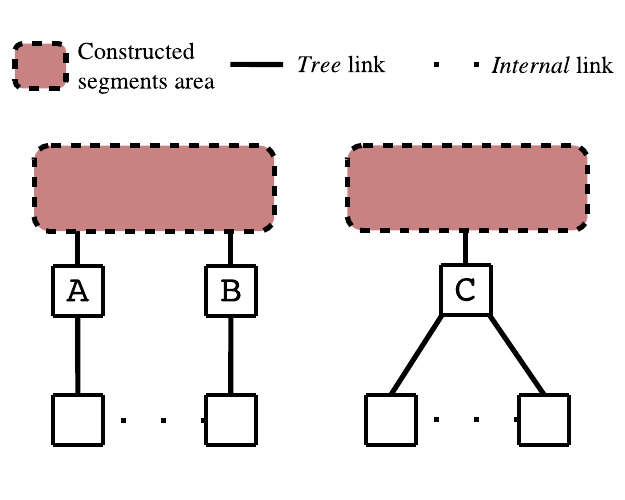}
\caption{Suitable \emph{internal} link (left) and an unsuitable
\emph{internal} link (right) to build new segment.}
\label{fig:segmentation:suitability}
\end{figure}

Fig. \ref{fig:segmentation:suitability} shows an example of suitable
(left) and unsuitable (right) links. Nodes $A$ and $B$ are connected but
not included within the constructed area, because they are not the left
link endpoints CAs, left link is considered suitable. $C$, for instance,
is not within the constructed area. Nevertheless, $C$ is CA of right
link endpoints. In consequence, right link is not suitable to build a
segment which connects to already constructed segments.

Initially, only the starting switch of the subnet (usually the MST root
switch) is considered to be within the constructed area, this allows to
build the starting segment.

Switches adjacent to the starting switch (i.e. switches connected to the
constructed area, but not included) are used to evaluate \emph{internal}
links suitability in the MST. \emph{Internal} links endpoints not having
any of these adjacent switches as CA are considered suitable to build
the starting segment. In other words, suitable \emph{internal} links
endpoints LCA must be the starting switch.

\begin{figure*}[!t]
    \centering
    \includegraphics[width=7in]{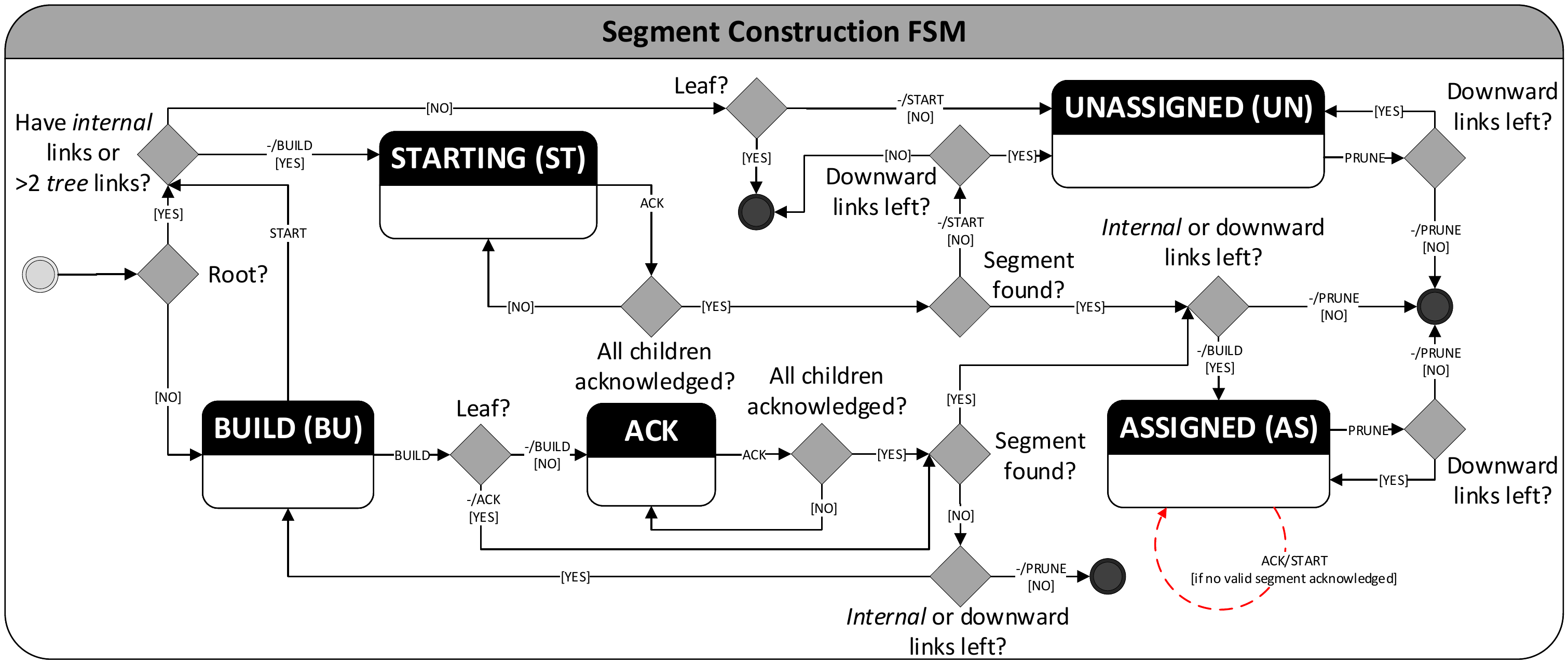}
    \caption{Segment construction FSM diagram.}
    \label{fig:segmentation-fsm-diagram}
\end{figure*}

Fig. \ref{fig:segmentation-fsm-diagram} shows the finite state machine
modeling the implemented behavior. This stage takes place after both
the MST construction and labeling stages have finished locally at each
switch.

The process of checking link suitability and segment building
is performed by BUILD and ACK messages, respectively. Nodes in
the constructed area (i.e. switches belonging to a constructed
segment) send empty BUILD messages through \emph{tree} links
towards switches not included in any segment ($A$ and $B$ in Fig.
\ref{fig:segmentation:suitability}). Directly connected switches to the
constructed area found downwards store their label in the BUILD message
and keep forwarding it.

Downward \emph{internal} links check their suitability using the
switch's label provided by the BUILD message (if any), ensuring that
this switch is not CA of the \emph{internal} link endpoints. The BUILD
message is forwarded unchanged by intermediate switches towards leaf
switches.

Upon arrival of BUILD messages, \emph{internal} links found (if
any) would be considered suitable either if\footnote{We assume that
\emph{internal} link endpoints are not CA of each other, otherwise the
\emph{internal} link would only be suitable to build a starting segment
within a new subnet.}:

\begin{enumerate}
    \item It has at least one of its endpoints connected directly to the
constructed area (BUILD message would be empty at this point).
    \item Otherwise, both endpoints must check that the stored switch
within the BUILD message is not a CA.
\end{enumerate}

For example, \emph{internal} link having endpoint switches $A$ and $B$
in Fig. \ref{fig:segmentation:suitability} would be suitable because
the first condition would be met (i.e. there is at least one endpoint
connected directly to the constructed area).

Leaf switches consume the BUILD message, and they send an ACK message
upwards. ACK messages carry the weight of the suitable \emph{internal}
link found (if any).

Along the path traveled by the ACK message, switches and links are
assigned to the segment indicated by the ACK message. Paths leading
to suitable \emph{internal} links may overlap. Nevertheless, segment
overlapping is not allowed by Rule \ref{rule:sr-1}. To avoid segment
overlapping, switches wait for arrival of ACK messages from all their
successors to be able to select the best link weight stored within all
the ACK messages. Finally, upon reception of successors' ACK messages,
an ACK message is sent upwards taking into account the weight of the
suitable \emph{internal} links found at this switch and the best weight
selected among its successors.

\begin{figure}[!t]
    \centering
    \includegraphics[width=2.0in]{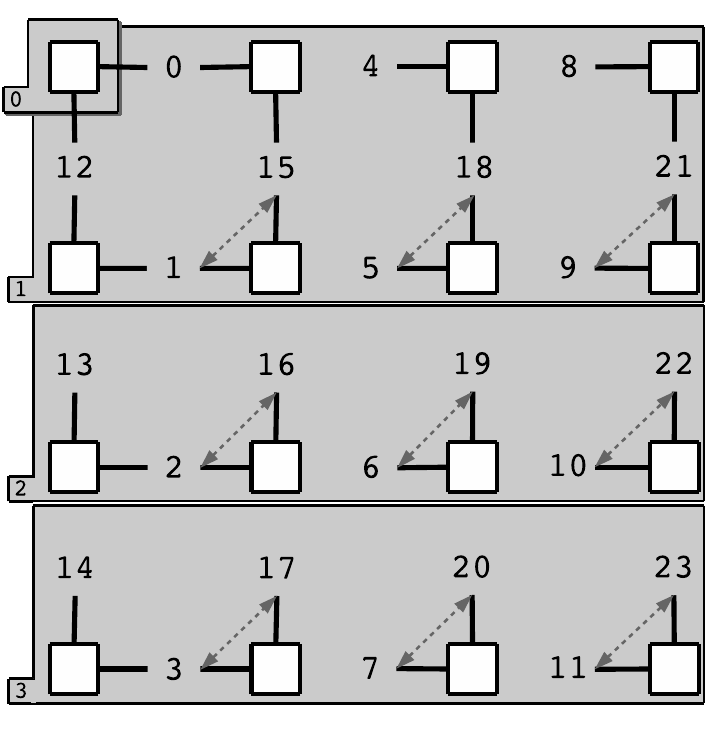}
    \caption{Segment construction example from an horizontal weight
distribution MST (see Fig. \ref{fig:4x4-tree:hfirst}). Grayed regions
show the three area expansions required to build nine segments covering
all \emph{internal} links. Routing restrictions within segments are also
shown using dashed arrows.}
    \label{fig:segmentation-example}
\end{figure}

Fig. \ref{fig:segmentation-example} shows segments constructed
for the tree shown in Fig. \ref{fig:4x4-tree:hfirst}. Initial
constructed area ($0$) started at root switch. Then, to build segments
covering the entire network, the constructed area required three
expansions ($1$, $2$ and $3$), each expansion adding new constructed
segments. \emph{Internal} links adding segments in a given area
expansion have their LCA endpoints within segments added by the
previous expansion. Note that segments do not overlap as lower weight
\emph{internal} links take precedence. The example consists of a single
subnet as no \emph{bridge} links exists.

Upon each segment construction, two routing restrictions (one at
each direction) must be placed within the segment to guarantee
deadlock-freedom upon routing packets through the network. Although the
routing restrictions could be placed anywhere within each segment, for
simplicity, we place a bidirectional routing restriction within the
switch with lower identifier connected through the \emph{internal} link
upon which the segment was built. However, a different criteria may be
used for routing restriction placement within segments.

During segment construction stage, some situations may arise which
require a more elaborated explanation. These cases are related to the
inability to build new segments:

\textbf{What if the starting segment cannot be built from the starting
switch?}  Assuming there exists at least one \emph{internal} link
downwards (otherwise, no segments would be built at all), the starting
switch withdraws from building the starting segment and sends a START
message downwards to its adjacent switches. Then it moves to the
\emph{UNASSIGNED} (UN) state. Downward switches receiving a START
message will be considered starting switches to build its starting
segment (i.e. they move to state \emph{STARTING} (ST)). For example,
switch $C$ at Fig. \ref{fig:segmentation:suitability} shows an example
where root switch (assuming the root switch is located within the
constructed area) would not be able to build the starting segment. Also,
starting switch outgoing links would be considered as bridge links
joining different subnets.

\textbf{What if no segment can be built from the constructed
segments area?}  This situation is observed in Fig.
\ref{fig:segmentation:suitability} for the unsuitable \emph{internal}
link at the right half of the figure. Similar to the previous
situation, we assume that at least one \emph{internal} link downwards
exists. Hence, a START message is sent downwards from the closest switch
within the constructed area. This message triggers the construction of
a new subnet while searching for the subnet's starting segment. In Fig.
\ref{fig:segmentation:suitability}, switch $C$ would receive the START
message from the constructed area. We can see that the \emph{internal}
link downwards would become a suitable candidate for building the
starting segment from $C$.  Therefore, a new subnet would be successfully
built.

\textbf{What if no \emph{internal} links are found?} No segments would
be built. This case may arise for networks (or network regions) where
all available links are \emph{tree} links. Therefore, by the MST
definition, no cycles would exist. Thus, there is no need to build any
segment nor place routing restrictions between different links
guaranteeing deadlock-freedom and preserving connectivity (provided by
\emph{tree} links within the MST).

Time complexity upper bound for the segment construction stage depends
on the number of area expansions required to include existing
\emph{internal} links into segments by having its LCA within the
constructed area. Also, for each area expansion, the longest path
explored downwards the tree looking for suitable \emph{internal} links,
will have a significant impact on the exploration time required. The
worst case scenario would be similar to the tree shown in Fig.
\ref{fig:4x4-tree:zigzag}, with the root located at one of the two
outermost switches. This gives as a result a MST with depth equal to the
amount of switches in the network. Then, time complexity upper bound can be
safely set as $O(N^2)$.

\section{Performance Evaluation}\label{sec:evaluation}

In this section we evaluate TDSR performance in terms of the execution
time required for the mechanism to converge to a valid solution in a
distributed fashion within different scenarios. Because of the different
aspects interacting with TDSR we present the simulation environment and
experiment setup details in the following sections.

\subsection{Simulation Environment}

We tested the TDSR mechanism on a cycle-accurate mesh network simulator with
the following characteristics:

\begin{itemize}
\item Links connecting adjacent switches can transmit at most one packet
per cycle.
\item Single cycle latency of packet transmission between adjacent switches.
\item TDSR control packets cannot be buffered, they must be processed upon
arrival.
\item TDSR control packet transmission takes precedence over any other packets
waiting to be transmitted through the same link.
\end{itemize}

This environment characterization aims to represent accurately the
systems targeted by TDSR such as many-core architectures featuring
networks-on-chip\cite{flich2018exploring}.

\subsection{Experiment Setup}

The experiment parameters are the network size, defective links
percentage over the total amount of links present in the network, and
link weight distribution.

Choosing different network sizes for the experiment allows us to test
TDSR scalability regarding the amount of switches present in the
network. This aspect also depends on the topology properties. Although
TDSR is a topology agnostic routing algorithm, for the sake of
brevity, only mesh topology is considered as it is a commonly
used topology for many-core architectures such as the U.T. Austin
Trips\cite{gratz2007chip}, Intel Teraflops\cite{vangal200780} and Tilera
TILE64\cite{wentzlaff2007chip}.

The fault model used in the experiment follows a uniform distribution
over links between adjacent switches. Due to transistor failures in
NoCs, switch's datapath components are likely to be affected by these
faults. Hence, driving individual links unusable\cite{fick2009vicis,
aisopos2011ariadne, lee2014brisk}.

Finally, the link weight distribution dictates the MST structure, which
greatly determines the area expansions required to assess link
suitability. Refer to Section \ref{sec:mst-stage} for a detailed
explanation.

Following, we present the different parameter configurations used
in this evaluation.

\begin{itemize}
\item Network size: $8\times8$, $16\times16$, $32\times32$ and $64\times64$.
\item Defective link rate: $0\%$ up to $45\%$.
\item Link weight distribution: \emph{center}, \emph{horizontal} and
    \emph{random}. \emph{Horizontal} and \emph{center} distributions are shown
in Fig. \ref{fig:4x4-tree:hfirst} and Fig. \ref{fig:4x4-tree:center}
respectively. On the other hand, the \emph{random} link weight distribution has
been generated by using a random sample from the available link weights without
replacement, ensuring each link gets a different unique weight. The random
sample follows a uniform distribution.
\end{itemize}

\subsection{Results}

The obtained results are shown according to the evaluation metric
defined previously, i.e. total execution time (in cycles) of the
algorithm.

\begin{figure*}[!t]
\centering
\subfloat[Horizontal link weight distribution]{
    \rotatebox{-270}{\makebox(0,0){\scriptsize \hspace{135pt} Time (cycles) \vspace{5pt}}}
    \includegraphics[width=0.31\textwidth]{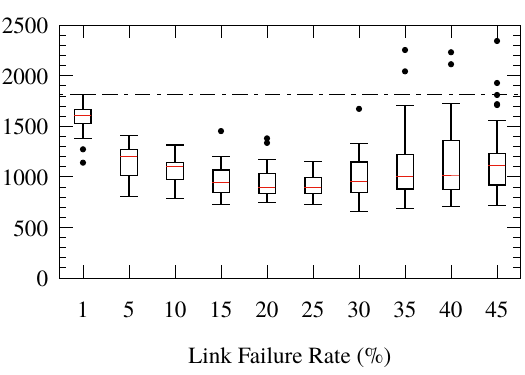}
    \label{fig:exec-time:16x16:horizontal}
}
\subfloat[Center link weight distribution]{
    \includegraphics[width=0.31\textwidth]{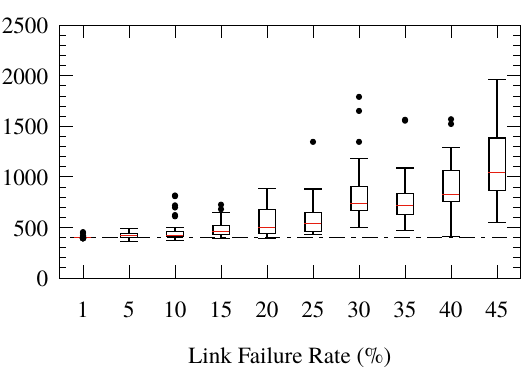}
    \label{fig:exec-time:16x16:center}
}
\subfloat[Random link weight distribution]{
    \includegraphics[width=0.31\textwidth]{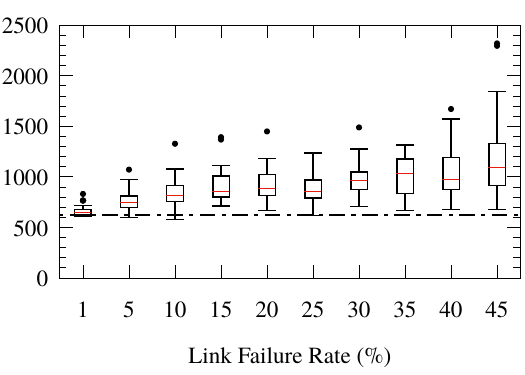}
    \label{fig:exec-time:16x16:random}
}
\caption{Execution time for 16x16 mesh configuration.}
\label{fig:exec-time:16x16}
\end{figure*}

For the sake of brevity, in this section we only show the
results for network size configuration $16\times16$ (Fig.
\ref{fig:exec-time:16x16}). Besides, we also provide network size
configurations $8\times8$, $32\times32$ and $64\times64$ results
in Appendix \ref{sec:results} (Figs. \ref{sup:fig:exec-time:8x8},
\ref{sup:fig:exec-time:32x32} and \ref{sup:fig:exec-time:64x64}
respectively).

For each network size, we have tested different link weight
distributions: \emph{center}, \emph{horizontal} and \emph{random}.
Finally, for each link weight distribution, total execution time
(vertical axes) is plotted by increasing the defective link rate
(horizontal axes).

Given a particular link weight distribution, TDSR operates in a
deterministic way. Hence, when no defective links are present in the
network, TDSR always converges to the same solution. This is represented
by an horizontal line in the results, showing the execution time of TDSR
with $0\%$ defective link rate. For defective link rates greater than
$0\%$, we run several executions of TDSR by randomly selecting which
links will be affected according to the defective rate such as to get
a representative sample. This data is plotted using boxplots (a.k.a.
Tukey boxplots\cite{frigge1989some}) in the figures.

Overall, results show that TDSR needs more cycles to converge if an
horizontal link weight distribution is chosen. This happens specially at
low defective link rates (see Fig. \ref{fig:exec-time:16x16:horizontal})
within each network size configuration. This is due to the depth of the
tree constructed at the MST construction stage, which in most cases is
greater than the tree obtained using \emph{center} and \emph{random}
link weight distributions.

Regarding \emph{center} and \emph{random} link weight distributions,
TDSR is able to converge to the solution keeping low the execution time
compared with the \emph{horizontal} distribution, specially at low
defective link rates.  Using the aforementioned distributions, TDSR
is less sensitive to an increase in defective link rates up to $20\%$
approximately.  Defective link rates above $20\%$ can result in a deeper
MST, which in turn raises the convergence time of the mechanism.

\emph{Random} link weight distribution usually needs more time to finish
than \emph{center} distribution. However, it is less sensitive to faults
distributed in a random uniform fashion. This means that the MST depth
in presence of defective links does not increase dramatically.

MST and segment construction stages described in sections
\ref{sec:mst-stage} and \ref{sec:segmentation-stage} are the most time
consuming stages of TDSR.  In the experiment performed, we have found
that the segment construction stage usually makes a larger impact
on the overall execution time of the mechanism. This stage is also
more sensitive than the MST construction stage to the link weight
distribution used.

\subsection{Analysis}

A brief discussion about the different aspects driving TDSR performance
under certain conditions is necessary.

\begin{figure*}[!t]
\centering
\subfloat[Horizontal link weight distribution]{
    \rotatebox{-270}{\makebox(0,0){\scriptsize \hspace{135pt} Time (cycles) \vspace{5pt}}}
    \includegraphics[width=0.31\textwidth]{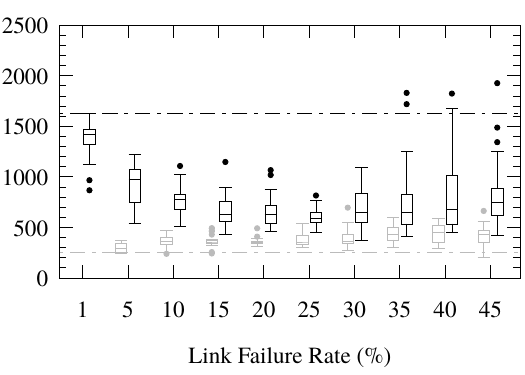}
    \label{fig:analysis:profile-horizontal}
}
\subfloat[Center link weight distribution]{
    \includegraphics[width=0.31\textwidth]{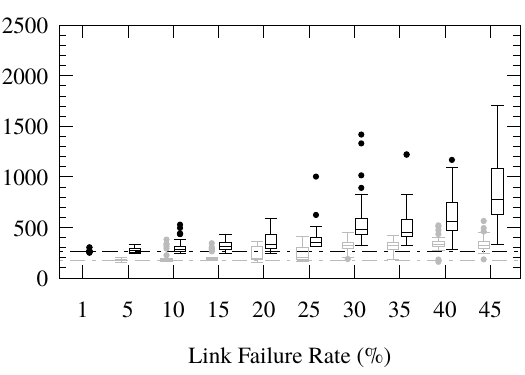}
    \label{fig:analysis:profile-center}
}
\subfloat[Random link weight distribution]{
    \includegraphics[width=0.31\textwidth]{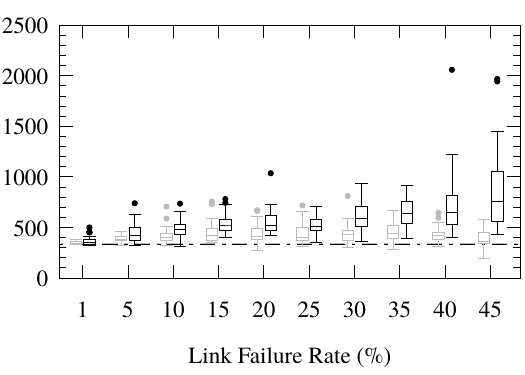}
    \label{fig:analysis:profile-random}
}
\caption{MST and Segment construction stages execution time for 16x16 mesh
configuration.}
\label{fig:analysis:profile}
\end{figure*}

MST and segment construction stages described in sections
\ref{sec:mst-stage} and \ref{sec:segmentation-stage} are the most time
consuming stages of TDSR.  In the experiment performed, we have found
that the segment construction stage usually makes a larger impact
on the overall execution time of the mechanism. This stage is also
more sensitive than the MST construction stage to the link weight
distribution used. Link weight distribution plays a critical role
regarding this aspect, as it determines the MST structure, root location
and in consequence, the MST depth. Fig. \ref{fig:analysis:profile} shows
the time each stage needed for the $16\times16$ mesh configuration.

\begin{figure}[!t]
\centering
\subfloat[]{
    \includegraphics[width=0.48\columnwidth]{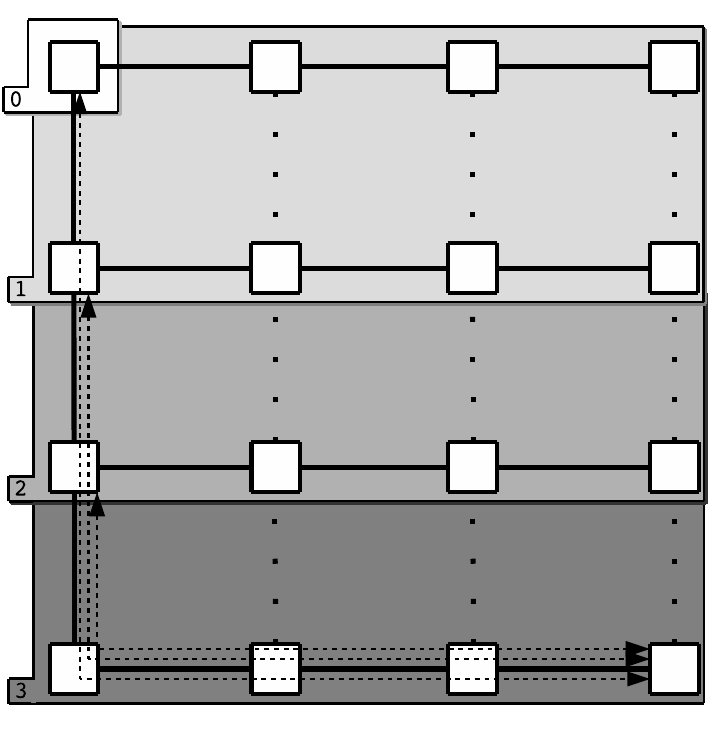}
    \label{fig:analysis:generations-horizontal}
}
\subfloat[]{
    \includegraphics[width=0.48\columnwidth]{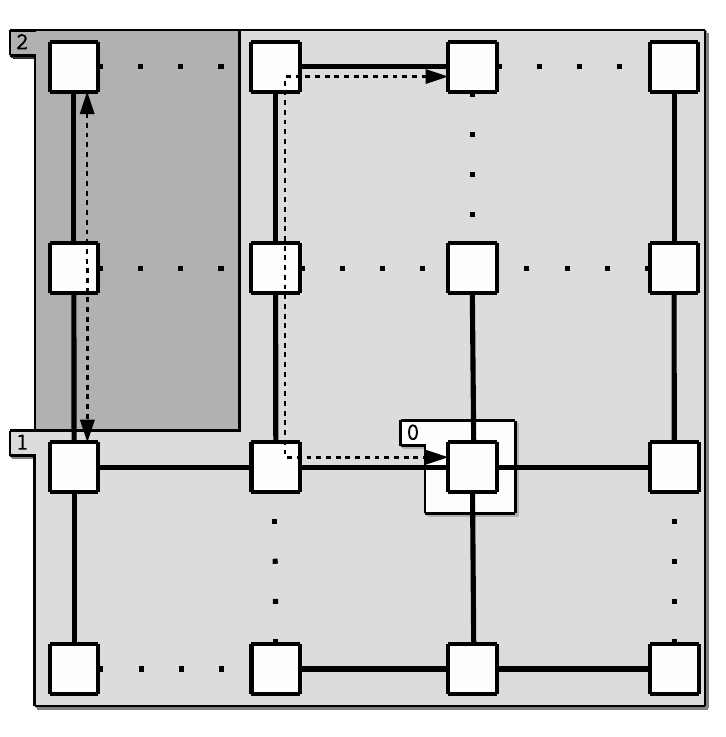}
    \label{fig:analysis:generations-center}
}
    \caption{Area expansions using horizontal (a) and center (b) link
weight distribution. Dashed arrows show longest paths explored at each
area expansion to construct segments.}
    \label{fig:analysis:generations-drawing}
\end{figure}

Fig. \ref{fig:analysis:generations-horizontal} shows the area
expansions required (grayed regions) to build all the segments using a
horizontal link weight distribution. Notice the large amount of switch
traversals made by TDSR in order to evaluate suitability of the distant
\emph{internal} link (located at the bottom right corner) per area
expansion (dashed arrows). Root location within the tree structure
generated by the horizontal link weight distribution has a significant
impact on the tree depth.

As a consequence, the execution time of TDSR decreases in presence of
defective links if using an horizontal link weight distribution (see Fig
\ref{fig:exec-time:16x16:horizontal}). This is due to the root location,
which in presence of defective links, it may be moved towards the MST
center.

On the other hand, if a center link weight distribution is configured, a
lower amount of area expansions are required to build the segments.
Also, the maximum distance traveled per area expansion is lowered as
Fig.  \ref{fig:analysis:generations-center} shows. In this case, due to
the initial centered location of the root with no defective links, the
presence of defective links may move the root away from its initial
location, thus, increasing the tree depth.

\begin{figure}[!t]
\centering
    \includegraphics[width=\columnwidth]{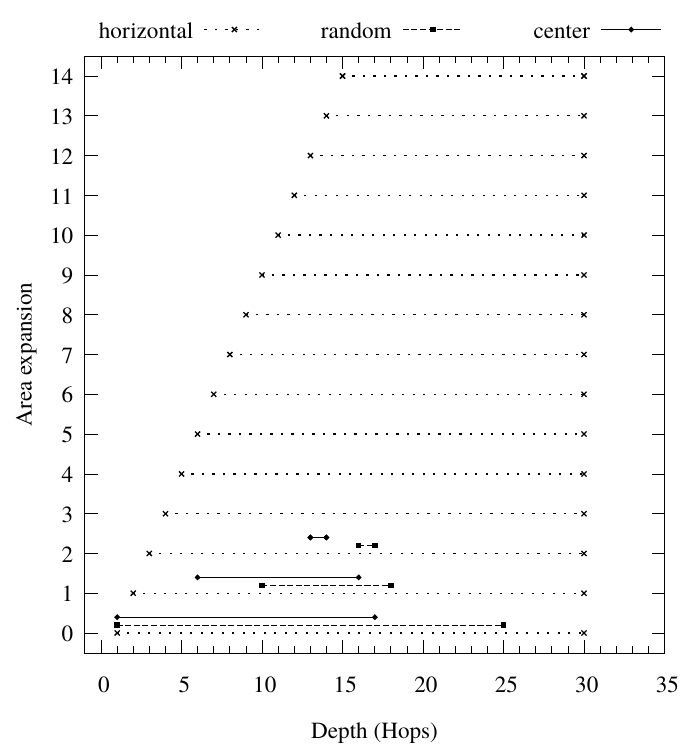}
    \caption{Distance traveled per area expansion in a 16x16 mesh using
different link weight distribution with no defective links.}
    \label{fig:analysis:generations}
\end{figure}

Finally, Fig. \ref{fig:analysis:generations} shows the distance
traveled per area expansion in a $16\times16$ mesh for each link weight
distribution with no defective links. The \emph{horizontal} distribution
needs up to 15 area expansions to build the segments with greater
distances traveled per area expansion. Meanwhile, \emph{center} and
\emph{random} distributions keep both factors low, although
\emph{random} distribution may travel slightly greater distances at some
area expansions.

Notice that the amount of area expansions required by the horizontal
distribution matches the amount of network rows minus one (starting
switch area expansion is omitted). This is showed also in Fig.
\ref{fig:analysis:generations-horizontal}.

\section{Conclusion}\label{sec:conclusion}

In this paper we have proposed a new SR-based\cite{mejia2006segment}
topology-agnostic deadlock-free distributed algorithm known as
\emph{Transparent Distributed Segment-based Routing (TDSR)}.

TDSR algorithm's main objective is to split the topology in subnets
which are further divided in segments in a distributed fashion. Thus,
allowing to devise a deadlock-free routing configuration by placing
routing restrictions within segments. Moreover, this objective must be
achieved guaranteeing connectivity among all switches within the same
network component.

Among TDSR key features is the flexibility of segment construction by
link weight distribution configuration. Also, depending on the link
weight distribution chosen, TDSR outperforms existing proposals such as
DiSR\cite{catania2014distributed} in multiple scenarios.

We have analyzed different aspects driving TDSR performance in terms of
execution time required to compute all segments.  We conclude that
choosing an appropriate link weight distribution has a significant
impact on the time required by TDSR. Link weight distributions reducing
the MST depth provide lower execution time due to its influence on the
exploration paths' length followed per area expansion, and the amount of
area expansions required. For instance, defective mesh networks may
benefit from a centered link weight distribution.

Future works will focus on setting routing restrictions within segments
according to some criteria such as network congestion, flow balancing,
etc. This will allow us to study the impact on network performance that
the different selection criteria of TDSR may produce.

Strategies to improve TDSR execution time should be further devised,
improving those aspects which have a significant impact on performance.

\section*{Acknowledgment}

This work has been jointly supported by the Spanish MINECO and European
Commission (FEDER funds) under the project RTI2018-098156-B-C52
(MINECO/FEDER), and by Junta de Comunidades de Castilla-La Mancha under
the project SBPLY/17/180501/000498. Juan-Jose Crespo is funded by the
Spanish MECD under national grant program (FPU) FPU15/03627. German
Maglione-Mathey is funded by the University of Castilla-La Mancha (UCLM)
with a pre-doctoral contract PREDUCLM16/29.

\appendices

\section{Segment-based Routing Pseudocode}\label{sec:sr}

Segment-based Routing (SR) \cite{mejia2006segment} is a
topology-agnostic routing algorithm aimed to provide a reasonable
path quality and fault-tolerance while keeping complexity low. SR is
considered a rule-driven routing algorithm \cite{flich2011survey}
whose main features and required resources can be summarized as follows:

\begin{enumerate}
    \item It does not guarantee shortest path computation by design.
    \item Virtual channels are not required.
    \item Deadlock freedom enforcement and path selection stages do not
rely on spanning tree computation.
\end{enumerate}

SR working principle is the partitioning of a topology into
subnets. Subnets in turn are made of disjoint segments where routing
restrictions are placed locally within each segment to break cycles,
thus guaranteeing deadlock freedom and connectivity within each
subnet. During the partitioning process, network links are visited at
most once to avoid the procedure to reach a deadlocked state.

A segment is defined as a list of interconnected switches and links. SR
pseudocode specification is shown in Algorithm~\ref{alg:sr}. For
simplicity, we use helper procedures as described below:

\begin{itemize}
    \item \emph{non\_visited\_switch}: It retreives a non visited switch
from either a set of Switches or a particular link.
    \item \emph{outgoing\_links}: Given a set of switches within a
subnet, it retrieves the outgoing links from that subnet (i.e. links
connecting a switch within the subnet, with a foreign switch which is
not contained in that subnet).
    \item \emph{path\_to\_segment}: It returns the set of switches and
links traversed starting at a link in a subnet to form a segment (i.e.
it must finish in an already computed segment within the same subnet).
\end{itemize}

\begin{algorithm}
\begin{algorithmic}[0]
\Require $S$ the set of Switches, $L$ the set of Links
\Procedure{SR}{$S$, $L$}
    \State $B \gets \emptyset$\Comment The set of bridge links 
    \State $sn \gets 0$\Comment Subnet counter
    \State $sg \gets 0$\Comment Segment counter
    \State $SG_s[sn][sg] \gets \emptyset$
    \Comment Segment switches per subnet
    \State $SG_l[sn][sg] \gets \emptyset$
    \Comment Segment links per subnet
    \While{$|SG_l| \neq |L| \land |SG_s| \neq |S|$}
        \If{$B = \emptyset$}
            \State $s \gets non\_visited\_switch(S)$
        \Else
            \State $s \gets non\_visited\_switch(link) : link \in B$
        \EndIf
        \State $SG_s[sn][sg] \gets SG_s[sn][sg] + s$
        \For{$link \in outgoing\_links(SG_s[sn])$}
            \State $S', L' \gets path\_to\_segment(link, SG_s[sn])$
            \If{$L' \neq \emptyset$}
                \State $SG_l[sn][sg] \gets SG_l[sn][sg] + L'$
                \State $SG_s[sn][sg] \gets SG_s[sn][sg] + S'$
                \State $sg \gets sg + 1$
            \Else
                \State $B \gets B + link$
            \EndIf
        \EndFor
        \State $sn \gets sn + 1$
        \State $sg \gets 0$
    \EndWhile
\EndProcedure
\end{algorithmic}
\caption{Segment-based Routing.}
\label{alg:sr}
\end{algorithm}

\section{Additional Results}\label{sec:results}

The obtained results are shown according to the evaluation metric
defined previously, i.e. total execution time (in cycles) of the
algorithm.

\begin{figure*}[!t]
\centering
\subfloat[Horizontal link weight distribution]{
    \rotatebox{-270}{\makebox(0,0){\scriptsize \hspace{135pt} Time (cycles) \vspace{5pt}}}
    \includegraphics[width=0.31\textwidth]{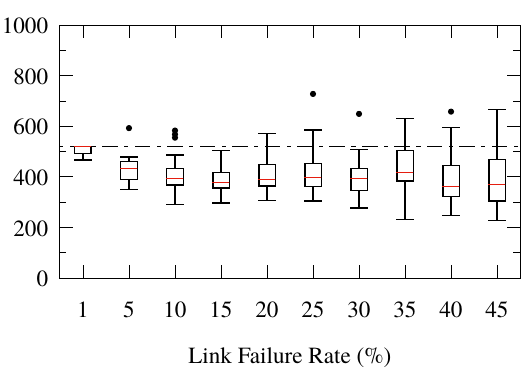}
    \label{sup:fig:exec-time:8x8:horizontal}
}
\subfloat[Center link weight distribution]{
    \includegraphics[width=0.31\textwidth]{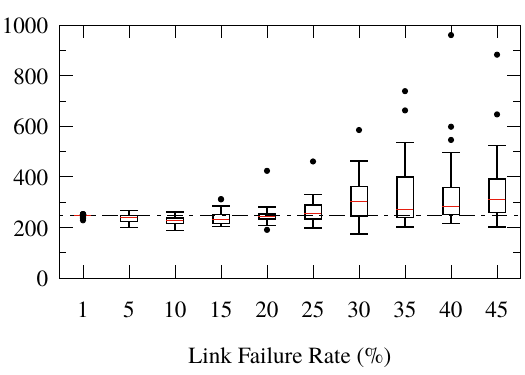}
    \label{sup:fig:exec-time:8x8:center}
}
\subfloat[Random link weight distribution]{
    \includegraphics[width=0.31\textwidth]{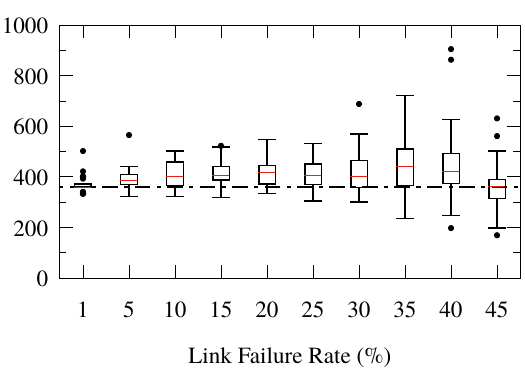}
    \label{sup:fig:exec-time:8x8:random}
}
\caption{Execution time for 8x8 mesh configuration.}
\label{sup:fig:exec-time:8x8}
\end{figure*}

\begin{figure*}[!t]
\centering
\subfloat[Horizontal link weight distribution]{
    \rotatebox{-270}{\makebox(0,0){\scriptsize \hspace{135pt} Time (cycles) \vspace{5pt}}}
    \includegraphics[width=0.31\textwidth]{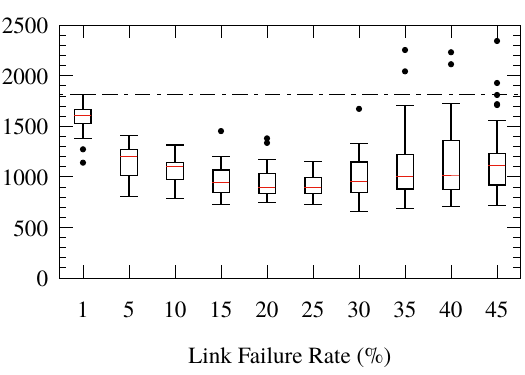}
    \label{sup:fig:exec-time:16x16:horizontal}
}
\subfloat[Center link weight distribution]{
    \includegraphics[width=0.31\textwidth]{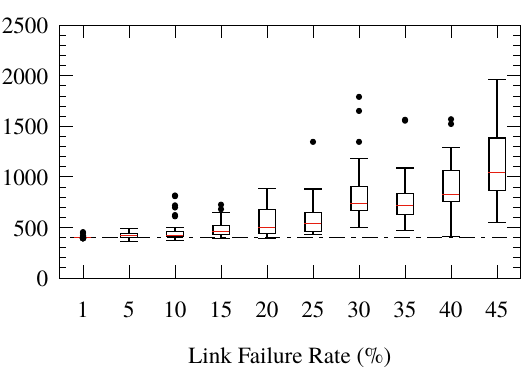}
    \label{sup:fig:exec-time:16x16:center}
}
\subfloat[Random link weight distribution]{
    \includegraphics[width=0.31\textwidth]{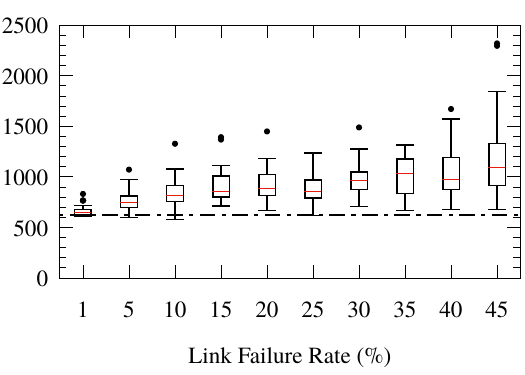}
    \label{sup:fig:exec-time:16x16:random}
}
\caption{Execution time for 16x16 mesh configuration.}
\label{sup:fig:exec-time:16x16}
\end{figure*}

\begin{figure*}[!t]
\centering
\subfloat[Horizontal link weight distribution]{
    \rotatebox{-270}{\makebox(0,0){\scriptsize \hspace{135pt} Time (cycles) \vspace{5pt}}}
    \includegraphics[width=0.31\textwidth]{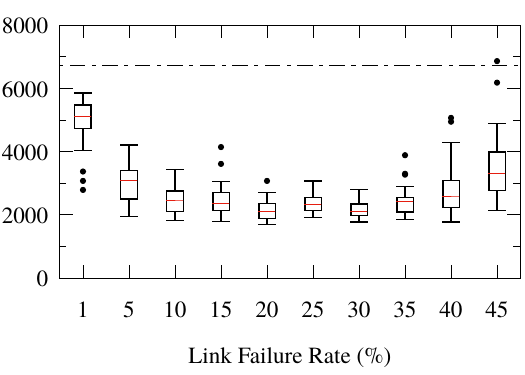}
    \label{sup:fig:exec-time:32x32:horizontal}
}
\subfloat[Center link weight distribution]{
    \includegraphics[width=0.31\textwidth]{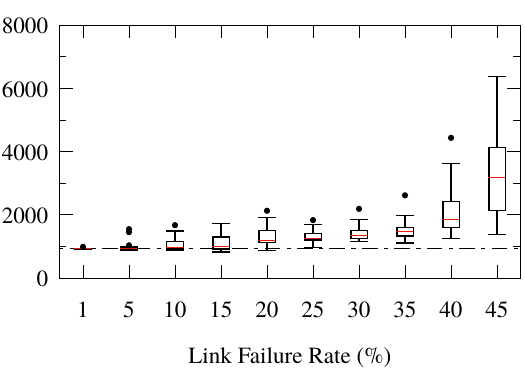}
    \label{sup:fig:exec-time:32x32:center}
}
\subfloat[Random link weight distribution]{
    \includegraphics[width=0.31\textwidth]{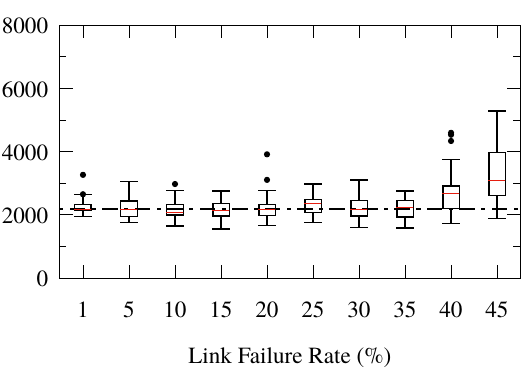}
    \label{sup:fig:exec-time:32x32:random}
}
\caption{Execution time for 32x32 mesh configuration.}
\label{sup:fig:exec-time:32x32}
\end{figure*}

\begin{figure*}[!t]
\centering
\subfloat[Horizontal link weight distribution]{
    \rotatebox{-270}{\makebox(0,0){\scriptsize \hspace{135pt} Time (cycles) \vspace{5pt}}}
    \includegraphics[width=0.31\textwidth]{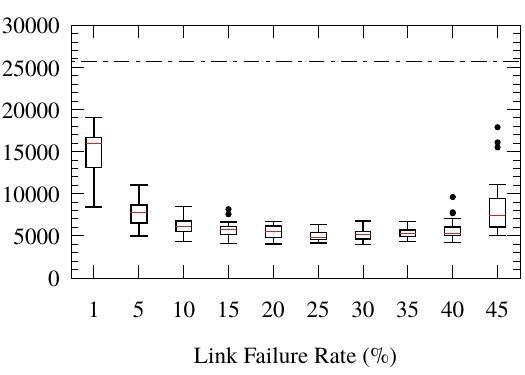}
    \label{sup:fig:exec-time:64x64:horizontal}
}
\subfloat[Center link weight distribution]{
    \includegraphics[width=0.31\textwidth]{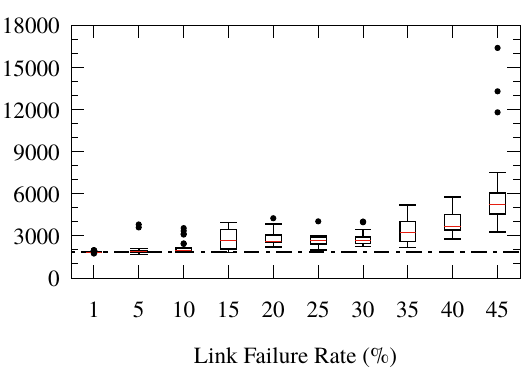}
    \label{sup:fig:exec-time:64x64:center}
}
\subfloat[Random link weight distribution]{
    \includegraphics[width=0.31\textwidth]{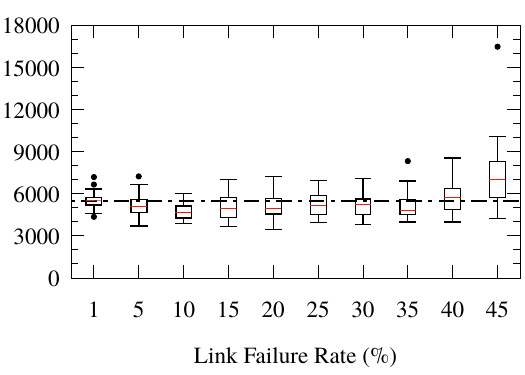}
    \label{sup:fig:exec-time:64x64:random}
}
\caption{Execution time for 64x64 mesh configuration.}
\label{sup:fig:exec-time:64x64}
\end{figure*}

Results are drawn for each network size configuration: $8\times8$ (Fig.
\ref{sup:fig:exec-time:8x8}), $16\times16$ (Fig. \ref{sup:fig:exec-time:16x16}),
$32\times32$ (Fig. \ref{sup:fig:exec-time:32x32}) and $64\times64$ (Fig.
\ref{sup:fig:exec-time:64x64}). For each network size, we have tested
different link weight distributions: \emph{center}, \emph{horizontal}
and \emph{random}.  Finally, for each link weight distribution, total
execution time (vertical axes) is plotted by increasing the defective
link rate (horizontal axes).

Given a particular link weight distribution, TDSR operates in a
deterministic way. Hence, when no defective links are present in the
network, TDSR always converges to the same solution. This is represented
by an horizontal line in the results, showing the execution time of TDSR
with $0\%$ defective link rate. For defective link rates greater than
$0\%$, we run several executions of TDSR by randomly selecting which
links will be affected according to the defective rate such as to get
a representative sample.

Overall, results show that TDSR needs more cycles to converge
if an horizontal link weight distribution is chosen. This
happens specially at low defective link rates (see Figs.
\ref{sup:fig:exec-time:8x8:horizontal}, \ref{sup:fig:exec-time:16x16:horizontal}
and \ref{sup:fig:exec-time:32x32:horizontal}) within each network size
configuration. This is due to the depth of the tree constructed at
the MST construction stage, which in most cases is greater than the
tree obtained using \emph{center} and \emph{random} link weight
distributions.

Regarding \emph{center} and \emph{random} link weight distributions,
TDSR is able to converge to the solution keeping low the execution
time compared with the \emph{horizontal} distribution, specially at
low defective link rates.  Using the aforementioned distributions,
TDSR is less sensitive to an increase in defective link rates up to
$20\%$ approximately.  Defective link rates above $20\%$ can result in a
deeper MST, which in turn raises the convergence time of the mechanism.
\emph{Random} link weight distribution usually needs more time to finish
than \emph{center} distribution. However, it is less sensitive to faults
distributed in a random uniform fashion (i.e. lower MST depth).

\clearpage
\clearpage

\bibliographystyle{IEEEtran}
\bibliography{IEEEabrv,bibliography.bib}

\begin{thebibliography}{10}
\providecommand{\url}[1]{#1}
\csname url@samestyle\endcsname
\providecommand{\newblock}{\relax}
\providecommand{\bibinfo}[2]{#2}
\providecommand{\BIBentrySTDinterwordspacing}{\spaceskip=0pt\relax}
\providecommand{\BIBentryALTinterwordstretchfactor}{4}
\providecommand{\BIBentryALTinterwordspacing}{\spaceskip=\fontdimen2\font plus
\BIBentryALTinterwordstretchfactor\fontdimen3\font minus
  \fontdimen4\font\relax}
\providecommand{\BIBforeignlanguage}[2]{{%
\expandafter\ifx\csname l@#1\endcsname\relax
\typeout{** WARNING: IEEEtran.bst: No hyphenation pattern has been}%
\typeout{** loaded for the language `#1'. Using the pattern for}%
\typeout{** the default language instead.}%
\else
\language=\csname l@#1\endcsname
\fi
#2}}
\providecommand{\BIBdecl}{\relax}
\BIBdecl

\bibitem{iba_2015}
\BIBentryALTinterwordspacing
\emph{The InfiniBand{\textregistered} Architecture Specification, Volume 1,
  Release 1.3}.\hskip 1em plus 0.5em minus 0.4em\relax The InfiniBand Trade
  Association, March 2015. [Online]. Available:
  \url{http://www.infinibandta.org}
\BIBentrySTDinterwordspacing

\bibitem{top500june2019}
``Top500 supercomputer list,'' \url{https://www.top500.org/lists/2019/06/},
  53rd edition, June 2019.

\bibitem{nordstrom1992using}
T.~Nordstr{\"o}m and B.~Svensson, ``Using and designing massively parallel
  computers for artificial neural networks,'' \emph{Journal of parallel and
  distributed computing}, vol.~14, no.~3, pp. 260--285, 1992.

\bibitem{de2007massively}
K.~De~Raedt, K.~Michielsen, H.~De~Raedt, B.~Trieu, G.~Arnold, M.~Richter,
  T.~Lippert, H.~Watanabe, and N.~Ito, ``Massively parallel quantum computer
  simulator,'' \emph{Computer Physics Communications}, vol. 176, no.~2, pp.
  121--136, 2007.

\bibitem{gratz2007chip}
P.~Gratz, C.~Kim, K.~Sankaralingam, H.~Hanson, P.~Shivakumar, S.~W. Keckler,
  and D.~Burger, ``On-chip interconnection networks of the trips chip,''
  \emph{IEEE Micro}, vol.~27, no.~5, pp. 41--50, 2007.

\bibitem{vangal200780}
S.~Vangal, J.~Howard, G.~Ruhl, S.~Dighe, H.~Wilson, J.~Tschanz, D.~Finan,
  P.~Iyer, A.~Singh, T.~Jacob \emph{et~al.}, ``An 80-tile 1.28 tflops
  network-on-chip in 65nm cmos,'' in \emph{2007 IEEE International Solid-State
  Circuits Conference. Digest of Technical Papers}.\hskip 1em plus 0.5em minus
  0.4em\relax IEEE, 2007, pp. 98--589.

\bibitem{wentzlaff2007chip}
D.~Wentzlaff, P.~Griffin, H.~Hoffmann, L.~Bao, B.~Edwards, C.~Ramey,
  M.~Mattina, C.-C. Miao, J.~F. Brown~III, and A.~Agarwal, ``On-chip
  interconnection architecture of the tile processor,'' \emph{IEEE micro},
  no.~5, pp. 15--31, 2007.

\bibitem{sodani2016knights}
A.~Sodani, R.~Gramunt, J.~Corbal, H.-S. Kim, K.~Vinod, S.~Chinthamani,
  S.~Hutsell, R.~Agarwal, and Y.-C. Liu, ``Knights landing: Second-generation
  intel xeon phi product,'' \emph{Ieee micro}, vol.~36, no.~2, pp. 34--46,
  2016.

\bibitem{fricker2019apparatus}
J.-P. Fricker and P.~Ferolito, ``Apparatus and method for multi-die
  interconnection,'' Jul.~30 2019, uS Patent App. 10/366,967.

\bibitem{esmaeilzadeh2011dark}
H.~Esmaeilzadeh, E.~Blem, R.~S. Amant, K.~Sankaralingam, and D.~Burger, ``Dark
  silicon and the end of multicore scaling,'' in \emph{2011 38th Annual
  international symposium on computer architecture (ISCA)}.\hskip 1em plus
  0.5em minus 0.4em\relax IEEE, 2011, pp. 365--376.

\bibitem{haselman2009future}
M.~Haselman and S.~Hauck, ``The future of integrated circuits: A survey of
  nanoelectronics,'' \emph{Proceedings of the IEEE}, vol.~98, no.~1, pp.
  11--38, 2009.

\bibitem{lysne2008efficient}
O.~Lysne, J.~M. Montanana, J.~Flich, J.~Duato, T.~M. Pinkston, and T.~Skeie,
  ``An efficient and deadlock-free network reconfiguration protocol,''
  \emph{IEEE Transactions on Computers}, vol.~57, no.~6, pp. 762--779, 2008.

\bibitem{lee2014brisk}
D.~Lee, R.~Parikh, and V.~Bertacco, ``Brisk and limited-impact noc routing
  reconfiguration,'' in \emph{2014 Design, Automation \& Test in Europe
  Conference \& Exhibition (DATE)}.\hskip 1em plus 0.5em minus 0.4em\relax
  IEEE, 2014, pp. 1--6.

\bibitem{mejia2006segment}
A.~Mejia, J.~Flich, J.~Duato, S.-A. Reinemo, and T.~Skeie, ``Segment-based
  routing: An efficient fault-tolerant routing algorithm for meshes and tori,''
  in \emph{Proceedings 20th IEEE International Parallel \& Distributed
  Processing Symposium}.\hskip 1em plus 0.5em minus 0.4em\relax IEEE, 2006, pp.
  10--pp.

\bibitem{gallager1983distributed}
R.~G. Gallager, P.~A. Humblet, and P.~M. Spira, ``A distributed algorithm for
  minimum-weight spanning trees,'' \emph{ACM Transactions on Programming
  Languages and systems (TOPLAS)}, vol.~5, no.~1, pp. 66--77, 1983.

\bibitem{nevsetvril2001otakar}
J.~Ne{\v{s}}et{\v{r}}il, E.~Milkov{\'a}, and H.~Ne{\v{s}}et{\v{r}}ilov{\'a},
  ``Otakar borůvka on minimum spanning tree problem translation of both the
  1926 papers, comments, history,'' \emph{Discrete mathematics}, vol. 233, no.
  1-3, pp. 3--36, 2001.

\bibitem{prim1957shortest}
R.~C. Prim, ``Shortest connection networks and some generalizations,''
  \emph{The Bell System Technical Journal}, vol.~36, no.~6, pp. 1389--1401,
  1957.

\bibitem{kruskal1956shortest}
J.~B. Kruskal, ``On the shortest spanning subtree of a graph and the traveling
  salesman problem,'' \emph{Proceedings of the American Mathematical society},
  vol.~7, no.~1, pp. 48--50, 1956.

\bibitem{chung1996parallel}
S.~Chung and A.~Condon, ``Parallel implementation of bouvka's minimum spanning
  tree algorithm,'' in \emph{Proceedings of International Conference on
  Parallel Processing}.\hskip 1em plus 0.5em minus 0.4em\relax IEEE, 1996, pp.
  302--308.

\bibitem{spira1977communication}
P.~Spira, ``Communication complexity of distributed minimum spanning tree
  algorithms,'' in \emph{Proceedings of the second Berkeley conference on
  distributed data management and computer networks}, 1977.

\bibitem{awerbuch1987optimal}
B.~Awerbuch, ``Optimal distributed algorithms for minimum weight spanning tree,
  counting, leader election, and related problems,'' in \emph{Proceedings of
  the nineteenth annual ACM symposium on Theory of computing}.\hskip 1em plus
  0.5em minus 0.4em\relax ACM, 1987, pp. 230--240.

\bibitem{garay1998sublinear}
J.~A. Garay, S.~Kutten, and D.~Peleg, ``A sublinear time distributed algorithm
  for minimum-weight spanning trees,'' \emph{SIAM Journal on Computing},
  vol.~27, no.~1, pp. 302--316, 1998.

\bibitem{santoro1985labelling}
N.~Santoro and R.~Khatib, ``Labelling and implicit routing in networks,''
  \emph{The computer journal}, vol.~28, no.~1, pp. 5--8, 1985.

\bibitem{flich2011survey}
J.~Flich, T.~Skeie, A.~Mejia, O.~Lysne, P.~Lopez, A.~Robles, J.~Duato,
  M.~Koibuchi, T.~Rokicki, and J.~C. Sancho, ``A survey and evaluation of
  topology-agnostic deterministic routing algorithms,'' \emph{IEEE Transactions
  on Parallel and Distributed Systems}, vol.~23, no.~3, pp. 405--425, 2011.

\bibitem{sancho2002effective}
J.~C. Sancho, A.~Robles, J.~Flich, P.~Lopez, and J.~Duato, ``Effective
  methodology for deadlock-free minimal routing in infiniband networks,'' in
  \emph{Proceedings International Conference on Parallel Processing}.\hskip 1em
  plus 0.5em minus 0.4em\relax IEEE, 2002, pp. 409--418.

\bibitem{skeie2002layered}
T.~Skeie, O.~Lysne, and I.~Theiss, ``Layered shortest path (lash) routing in
  irregular system area networks,'' in \emph{ipdps}.\hskip 1em plus 0.5em minus
  0.4em\relax Citeseer, 2002, p. 0162.

\bibitem{skeie2004lash}
T.~Skeie, O.~Lysne, J.~Flich, P.~Lopez, A.~Robles, and J.~Duato, ``Lash-tor: A
  generic transition-oriented routing algorithm,'' in \emph{Proceedings. Tenth
  International Conference on Parallel and Distributed Systems, 2004. ICPADS
  2004.}\hskip 1em plus 0.5em minus 0.4em\relax IEEE, 2004, pp. 595--604.

\bibitem{schroeder1991autonet}
M.~D. Schroeder, A.~D. Birrell, M.~Burrows, H.~Murray, R.~M. Needham, T.~L.
  Rodeheffer, E.~H. Satterthwaite, and C.~P. Thacker, ``Autonet: A high-speed,
  self-configuring local area network using point-to-point links,'' \emph{IEEE
  Journal on Selected Areas in Communications}, vol.~9, no.~8, pp. 1318--1335,
  1991.

\bibitem{sancho2000flexible}
J.~C. Sancho, A.~Robles, and J.~Duato, ``A flexible routing scheme for networks
  of workstations,'' in \emph{International Symposium on High Performance
  Computing}.\hskip 1em plus 0.5em minus 0.4em\relax Springer, 2000, pp.
  260--267.

\bibitem{koibuchi2001turn}
M.~Koibuchi, A.~Funahashi, A.~Jouraku, and H.~Amano, ``L-turn routing: An
  adaptive routing in irregular networks,'' in \emph{International Conference
  on Parallel Processing, 2001.}\hskip 1em plus 0.5em minus 0.4em\relax IEEE,
  2001, pp. 383--392.

\bibitem{zhou2012tree}
J.~Zhou and Y.-C. Chung, ``Tree-turn routing: an efficient deadlock-free
  routing algorithm for irregular networks,'' \emph{The Journal of
  Supercomputing}, vol.~59, no.~2, pp. 882--900, 2012.

\bibitem{catania2014distributed}
V.~Catania, A.~Mineo, S.~Monteleone, and D.~Patti, ``Distributed topology
  discovery in self-assembled nano network-on-chip,'' \emph{Computers \&
  Electrical Engineering}, vol.~40, no.~8, pp. 292--306, 2014.

\bibitem{anjan1995efficient}
K.~Anjan and T.~M. Pinkston, ``An efficient, fully adaptive deadlock recovery
  scheme: Disha,'' in \emph{ACM SIGARCH Computer Architecture News}, vol.~23,
  no.~2.\hskip 1em plus 0.5em minus 0.4em\relax ACM, 1995, pp. 201--210.

\bibitem{ramrakhyani2017static}
A.~Ramrakhyani and T.~Krishna, ``Static bubble: A framework for deadlock-free
  irregular on-chip topologies,'' in \emph{2017 IEEE International Symposium on
  High Performance Computer Architecture (HPCA)}.\hskip 1em plus 0.5em minus
  0.4em\relax IEEE, 2017, pp. 253--264.

\bibitem{ramrakhyani2018synchronized}
A.~Ramrakhyani, P.~V. Gratz, and T.~Krishna, ``Synchronized progress in
  interconnection networks (spin): A new theory for deadlock freedom,'' in
  \emph{2018 ACM/IEEE 45th Annual International Symposium on Computer
  Architecture (ISCA)}.\hskip 1em plus 0.5em minus 0.4em\relax IEEE, 2018, pp.
  699--711.

\bibitem{flich2018exploring}
J.~Flich, G.~Agosta, P.~Ampletzer, D.~A. Alonso, C.~Brandolese, E.~Cappe,
  A.~Cilardo, L.~Dragi{\'c}, A.~Dray, A.~Duspara \emph{et~al.}, ``Exploring
  manycore architectures for next-generation hpc systems through the mango
  approach,'' \emph{Microprocessors and Microsystems}, vol.~61, pp. 154--170,
  2018.

\bibitem{fick2009vicis}
D.~Fick, A.~DeOrio, J.~Hu, V.~Bertacco, D.~Blaauw, and D.~Sylvester, ``Vicis: a
  reliable network for unreliable silicon,'' in \emph{Proceedings of the 46th
  Annual Design Automation Conference}.\hskip 1em plus 0.5em minus 0.4em\relax
  ACM, 2009, pp. 812--817.

\bibitem{aisopos2011ariadne}
K.~Aisopos, A.~DeOrio, L.-S. Peh, and V.~Bertacco, ``Ariadne: Agnostic
  reconfiguration in a disconnected network environment,'' in \emph{2011
  International Conference on Parallel Architectures and Compilation
  Techniques}.\hskip 1em plus 0.5em minus 0.4em\relax IEEE, 2011, pp. 298--309.

\bibitem{frigge1989some}
M.~Frigge, D.~C. Hoaglin, and B.~Iglewicz, ``Some implementations of the
  boxplot,'' \emph{The American Statistician}, vol.~43, no.~1, pp. 50--54,
  1989.

\end{thebibliography}

\end{document}